\documentclass{article}

\usepackage{arxiv}

\usepackage[utf8]{inputenc} 
\usepackage[T1]{fontenc}    
\usepackage{hyperref}       
\usepackage{url}            
\usepackage{booktabs}       
\usepackage{amsfonts}       
\usepackage{nicefrac}       
\usepackage{microtype}      
\usepackage{lipsum}		
\usepackage{graphicx}
\usepackage[numbers]{natbib}
\usepackage{doi}
\usepackage{xspace}
\usepackage{multirow}
\usepackage{amsmath}

\def\eg{\emph{e.g.,}\xspace}

\def\ie{\emph{i.e.,}\xspace}
\def\etal{\emph{et al.}\xspace}

\title{A Meta-Analysis of LLM Effects on Students across Qualification, Socialisation, and Subjectification}

\author{%
Jiayu Huang$^{1,*}$,
Ruoxin Ritter Wang$^{2,*}$,
Jen\mbox{-}Hao Liu$^{1}$,
Boming Xia$^{3}$,
Yue Huang$^{4}$,
Ruoxi Sun$^{4}$,\\
\textbf{Jason (Minhui) Xue}$^{3,4}$,
\textbf{Jinan Zou}$^{1,\dagger}$ \\
\\
$^{1}$Australian Institute for Machine Learning, The University of Adelaide \\
$^{2}$Centre for Research in Educational and Social Inclusion, Education Futures, University of South Australia \\
$^{3}$Responsible AI Research (RAIR) Centre, The University of Adelaide \\
$^{4}$CSIRO's Data61 \\
\\
* Equal contribution \quad $\dagger$~Corresponding author
}

\renewcommand{\shorttitle}{A Meta-Analysis of LLM Effects on Students}

\begin{document}
\maketitle

\begin{abstract}
	Large language models (LLMs) are increasingly positioned as solutions for education, yet evaluations often reduce their impact to narrow performance metrics. This paper reframes the question by asking ``what kind of impact should LLMs have in education?'' Drawing on Biesta's tripartite account of good education: qualification, socialisation, and subjectification, we present a meta-analysis of 133 experimental and quasi-experimental studies ($k$ = 188). Overall, the impact of LLMs on student learning is positive but uneven. Strong effects emerge in qualification, particularly when LLMs function as tutors in sustained interventions. Socialisation outcomes appear more variable, concentrated in sustained, reflective interventions. Subjectification, linked to autonomy and learner development, remains fragile, with improvements confined to small-scale, long-term studies. This purpose-level view highlights design as the decisive factor: without scaffolds for participation and agency, LLMs privilege what is easiest to measure while neglecting broader aims of education. For HCI and education, the issue is not just whether LLMs work, but what futures they enable or foreclose.
\end{abstract}

\keywords{Large Language Models (LLMs) \and Meta-analysis \and Qualification \and Socialisation \and Subjectification \and Education }

\section{Introduction}
Large language models (LLMs) are entering classrooms at an unprecedented pace, serving as tutors, writing partners, and coding assistants~\cite{malinka2023educational,mcdonald2025generative}. Their adoption is not only an educational shift but also a human–computer interaction (HCI) challenge: these systems reshape how learners engage with knowledge, peers, and technology. Because the design of AI-mediated platforms will determine whether LLMs foster agency, collaboration, and long-term growth or instead reduce learning to quick answers and short-term gains, their rapid rise has sparked intense debate~\cite{kasneci2023chatgpt,bai2023future}. Advocates highlight opportunities to democratise knowledge and lighten teacher workload~\cite{kasneci2023chatgpt}, while critics warn of diminished critical thinking, weakened academic integrity, and the over-automation of judgement~\cite{rudolph2023chatgpt}.

Amid these debates, empirical studies highlight both the potential and the pitfalls of LLMs in education. Experimental studies point to gains in achievement, higher-order thinking, and engagement~\cite{lehmann2024ai,chen2024learning}, and randomised field trials show improved short-term performance when LLMs are used with structured prompts~\cite{liu2025can}. However, other findings caution that unguided or answer-provision use can harm long-term learning~\cite{etkin2025differential,bastani2025generative}, and prior work warns that over-reliance may undermine creativity and learner autonomy~\cite{kreijkes2025effects,zhai2024effects}. These tensions reflect wider philosophical and policy divides: governments oscillate between promoting responsible use~\cite{vidal2023emerging,holmes2023guidance} and imposing school bans~\cite{vincent2023new}, while teachers and students juggle new opportunities with fears of inequity and disempowerment.

Most existing evidence, however, is narrowly framed. Current meta-analyses report mainly positive effects~\cite{deng2025does,heung2025chatgpt,zhu2025exploring}, but almost all treat educational success as improved test scores and skills. As noted by Selwyn~\cite{selwyn2016technology} and Williamson \& Piattoeva~\cite{williamson2022education}, this ``age of measurement'' privileges what is easiest to quantify while neglecting what is hardest to value. This leaves three gaps: (i) it zooms in on easily measured performance while underreporting interactional participation and learner agency; (ii) it offers little theoretical basis for asking what kinds of education LLMs enable (\ie educational purposes); and (iii) it limits empirical coverage (\eg narrow subjects, higher education, short interventions) and leaves open how effects vary across contexts and over longer durations. This requires a conceptual shift from measuring isolated skills to evaluating holistic educational aims.

We address these gaps using Biesta's tripartite framework of education: qualification, socialisation, and subjectification~\cite{biesta2009good,biesta2015good,biesta2021world}. This enables us to ask not just whether LLMs raise performance, but also how they shape students' social participation and development as autonomous subjects. We conduct a meta-analysis of 133 experimental and quasi-experimental studies ($k = 188$) following PRISMA 2020 guidelines~\cite{page2021prisma, liberati2009prisma}, reclassify outcomes into Biesta's three purposes, and test how design factors (\eg subject, duration, instructional design) moderate effects. To address the coverage gap, we explicitly code context and design (role, duration, learning strategy, level, subject, region), quantify where evidence is concentrated, and estimate context-specific effects rather than pooling brief trials. This provides the first purpose-level synthesis of LLMs' educational impact, aiming to inform both educational research and HCI debates on how interactive technologies affect human agency, identity, and community.

This study is guided by three research questions: 
\begin{itemize}
    \item \textbf{RQ1:} What are the characteristics of student-facing LLM interventions reported to date?
    \item \textbf{RQ2:} What are the aggregate effects of LLM interventions on Biesta's three purposes: qualification, socialisation, and subjectification?
    \item \textbf{RQ3:} Which intervention characteristics (\eg LLM role, duration, subject area, educational level, learning strategy) are associated with effect variation within each purpose?
\end{itemize}

The remainder proceeds as follows: \S\ref{sec_related_work} reviews existing scholarship on LLMs in education and situates our work within broader debates on educational technology and its purposes. \S\ref{sec_theoretical_framework} presents the theoretical framework (Biesta) and its interpretive role. \S\ref{sec_methodology} describes the methodology of our meta-analysis, including data collection, coding, and statistical procedures. \S\ref{sec_results} presents the findings, first by mapping the descriptive landscape, then by reporting overall effects, and finally by examining key moderators. \S\ref{sec:discussion} then interprets these findings in relation to each research question and the broader literature. Building on this analysis, \S\ref{sec_implication} outlines their implications for theory, method, practice, and policy. Finally, \S\ref{sec_conclusion} concludes by reflecting on what these results suggest for the future of education in an age of LLMs.

\section{Related Work}
\label{sec_related_work}

\subsection{Student-Facing LLMs: Empirical Baseline}

Work on student-facing LLM uses (\eg writing support~\cite{meyer2024using, han2024exploring, guo2024resist}, problem-solving~\cite{urban2024chatgpt, sun2024would}, and exam preparation~\cite{yusof2025chatgpt, al2025exploring}) has expanded rapidly since late 2022. Across language learning~\cite{deep2025chatgpt,baskara2023exploring,yu2025can}, computer science~\cite{chen2023gptutor, sun2024would, husain2024potentials}, and health education~\cite{mackey2024evaluating, aster2025chatgpt}, studies report gains in achievement and higher-order thinking, with effects that vary by setting and design~\cite{lehmann2024ai,chen2024learning}.
Field and classroom experiments also identify boundary conditions. Benefits are larger when LLMs are embedded in structured, guided workflows that position the system as a tutor or partner, while unguided or answer-provision use can produce neutral or negative longer-term outcomes~\cite{vanzo2024gpt, bastani2025generative}. These patterns suggest that design choices such as LLM role and duration are consequential, and that ``LLM use'' should not be treated as a single treatment. We focus the remainder of this section on closely related reviews and meta-analyses, since our contribution is a synthesis at that level and the next section provides the normative frame used to interpret outcomes.

\subsection{Prior Syntheses: Coverage and Methods}

Our map of the ten most closely related meta-analyses (Table~\ref{tab:relatedwork}) reveals three regularities.

\begin{table}[t]
\centering
\caption{Overview of Related Meta-Analyses}
\label{tab:relatedwork}
\resizebox{\textwidth}{!}{%
\begin{tabular}{p{3.2cm}p{1.2cm}p{0.6cm}p{2.8cm}p{1.6cm}p{4.5cm}p{2.5cm}}
\toprule
\textbf{Author (Year)} & \textbf{Period} & \textbf{N} & \textbf{Education Level} & \textbf{Scope} & \textbf{Outcome Focus} & \textbf{Method} \\ \midrule

\textbf{Ours} & 2022– 2025.3 & \textbf{133} & \textbf{Kindergarten}, K12, Higher Education & broader LLMs & \textbf{Qualification} (e.g., academic performance), \textbf{Socialisation} (e.g., collaboration engagement), \textbf{Subjectification} (e.g., self-efficacy) & Meta-analyses + Moderator analyses \\

Deng~\etal (2025)~\cite{deng2025does} & 2022.11– 2024.8 & 69 & K12, Higher Education & ChatGPT only & Qualification (Academic performance, higher-order thinking), Subjectification (affective-motivational states, self-efficacy) & Meta-analyses + Moderator analyses \\ 

Sun \& Zhou (2024)~\cite{sun2024does} & 2022.11– 2024 & 28 & Higher Education & broader LLMs & Qualification (Academic achievement) & Meta-analyses + Moderator analyses \\ 

Liu~\etal (2025)~\cite{liu2025effects} & 2022– 2024.8 & 49 & K12, Higher Education & broader LLMs & Qualification (Learning achievement), Subjectification (learning motivation) & Meta-analyses + Moderator analyses \\ 

Hu~\etal (2025)~\cite{hu2025evaluating} & 2022– 2024.5 & 21 & K12, Higher Education & broader LLMs &  Qualification (Bloom's taxonomy: cognitive, behavioural), Subjectification (affective) & Meta-analyses + Moderator analyses \\ 

Zhu~\etal (2025)~\cite{zhu2025exploring} & 2020– 2024.8 & 28 & K12, Higher Education & broader LLMs & Qualification (Physical, intellectual outcomes), Subjectification (social-emotional outcomes) & Meta-analyses + Moderator analyses \\ 

Qu~\etal (2025)~\cite{qu2025generative} & 2023.1– 2024.9 & 18 & Higher Education & broader LLMs & Qualification (Cognitive outcomes in Bloom's taxonomy) & Meta-analyses + Regression\\ 

Heung \& Chiu (2025)~\cite{heung2025chatgpt} & 2021– 2024.8 & 17 & K12, Higher Education & ChatGPT only &  Subjectification (learning engagement) & Meta-analyses \\ 

Wang \& Fan (2025)~\cite{wang2025effect} & 2022.11– 2025.2 & 51 & K12, Higher Education & ChatGPT only & Qualification (Learning performance, higher-order thinking) & Meta-analyses + Moderator analyses \\ 

Liu~\etal (2025)~\cite{liu2025impact} & 2022.11– 2025.3 & 37 & K12, Higher Education & ChatGPT only & Qualification (Academic achievement) & Meta-analyses + Moderator analyses \\ 

Xia~\etal (2025)~\cite{xia2025impact} & Not reported & 24 & Higher Education & broader LLMs & Qualification (cognitive skill in Bloom's taxonomy), Subjectification (non-cognitive skills) & Meta-analyses + Moderator analyses + Pattern mining \\ 

\bottomrule
\end{tabular}
}
\end{table}

The dominant focus of the syntheses is on performance outcomes. In our sample, nine of the ten syntheses focus exclusively on qualification outcomes, such as knowledge and skill performance. Three of these organize their findings using Bloom's Taxonomy~\cite{krathwohl2002revision}, a scheme useful for grading cognitive demand. However, Bloom-based classifications assess the the type or level of cognition rather than the purposes of education, largely overlooking crucial dimensions like social participation and learner autonomy. Consequently, no review in our set synthesizes interactional outcomes that align with socialisation, and only one includes proxies for self-efficacy that speak to subjectification. The dominant evidence base, therefore, reflects a narrow focus on task-based performance, with much thinner coverage of student participation or agency.

The reviews vary in both population and scope. Seven cover both K–12 and higher education, while only three focus exclusively on higher education. The technological scope is also mixed, with four papers restricted to ChatGPT and six including broader LLMs. Most time windows span 2022–2024, though two extend into 2025 and one does not report a window. The number of included studies per review varies widely~\cite{heung2025chatgpt, deng2025does}, which cautions against over-generalisation.

The analytic strategies used in these reviews are limited and uneven. Eight of the ten reviews conduct moderator analyses, and one additionally applies association-pattern mining alongside meta-analytic models~\cite{xia2025impact}. Despite this analytic activity, most syntheses pool mixed outcome types and interpret effects primarily as evidence of “what works” for performance. Only a few attempt to link design conditions to differential impacts on outcomes beyond performance.

These regularities motivate the present study in two ways. First, the field offers a partial account of educational impact by equating success with performance. Second, existing syntheses seldom link design conditions to which educational purposes benefit.

\subsection{Design Moderating Factors in Prior Work}

Across recent meta-analyses reviews, several study and design factors recur as sources of heterogeneity in student-facing LLM deployments \cite{deng2025does,liu2025impact,heung2025chatgpt,xia2025impact}. Frequently discussed are design conditions such as LLM role (tutor or partner vs. learning tool), intervention duration (one-shot vs. multi-week exposure), and the learning strategy embedded in the activity (for example, problem-based, project-based, reflective). Context factors also appear often, including educational level (K-12 vs. higher education), subject category (language, computer science or medical), and sample size. Narrative syntheses often note stronger gains for dialogic tutor or partner deployments sustained over time, with more mixed results for brief, tool-like use \cite{deng2025does, liu2025impact}. Classroom and field experiments report similar boundary conditions when comparing structured and scaffolded workflows to unguided answer provision \cite{bastani2025generative}.

What is less developed in prior syntheses is an explicit connection between these conditions and outcomes beyond performance. When moderator analyses are reported, they typically pool mixed outcomes into a single effect size (for example, aggregating cognition, behavior, and affection learning outcome)~\cite{heung2025chatgpt, zhu2025exploring, hu2025evaluating}. This makes it difficult to see whether factors like role or duration differentially benefit collaboration and participation (socialisation) or autonomy and self-regulation (subjectification).

\textit{Positioning relative to our contributions.} We draw from five major databases and cover from 2022 to 2025.3, synthesizing 133 experimental or quasi-experimental studies. This is larger than most reviews in our set (median 28, range 17~\cite{heung2025chatgpt} to 69~\cite{deng2025does}). Our corpus spans higher education and K-12 and includes a small kindergarten subset. We track model families, including ChatGPT-class systems and others. Rather than collapsing to a single effectiveness construct, we organize outcomes using Biesta's purposes of qualification, socialisation, and subjectification, addressing the qualification-centric emphasis observed across the 10 reviews (Section~\ref{sec_related_work}). We then relate the recurring conditions identified in the literature to which educational purposes are advanced, instead of treating LLM use as a single undifferentiated treatment.

\section{Theoretical Framework}
\label{sec_theoretical_framework}

Evaluations of educational technologies have often emphasised short-term performance indicators such as test scores, task accuracy, and assignment grades \cite{selwyn2016technology, williamson2022education}. Policy guidance likewise foregrounds measurable attainment and assessment alignment \cite{holmes2023guidance, vidal2023emerging}. These indicators are useful, but a performance-first framing narrows what counts as impact. Following Biesta, questions of purpose should come first: ``only when we have a meaningful and justifiable answer to what our educational endeavors are for can we decide about content and relationships'' \cite{biesta2024taking}. We therefore interpret effects through Biesta's account of education's three purposes \cite{biesta2009good, biesta2015good, biesta2021world}.

\begin{itemize}
  \item \textbf{Qualification}: the acquisition of knowledge, skills, and dispositions that enable learners to ``do something.'' This includes not only vocational and academic competencies but also broader life skills. Although education systems increasingly equate qualification with credentials for labour market success, Biesta stresses that it encompasses a wider set of abilities and responsibilities. 
  \item \textbf{Socialisation}: the processes through which individuals become part of social, cultural, or political communities. Education socialises learners by transmitting norms, values, and professional standards, but also by shaping their implicit understandings of belonging and participation.
  \item \textbf{Subjectification}: the formation of learners as autonomous and responsible beings. It is about enabling individuals to take up their own positions, make choices, and respond to the world not merely as objects of external expectations but as subjects of their own lives. Subjectification does not mean unrestricted freedom; rather, it emphasises reflective engagement and the negotiation of limits.
\end{itemize}

These purposes are interdependent and often in tension; privileging one can erode the others. As Biesta cautions, ``one-sidedness always comes at a price'' \cite{biesta2015good}. In LLM-mediated settings, for example, designs that emphasise answer provision can raise qualification metrics while dampening participation or autonomy.

In the analyses that follow, we read outcomes at the level of purposes rather than collapsing them into a single effectiveness score. 

This lens is pertinent to HCI because design choices in human-AI systems (\eg role, duration) allocate control and shape participation and identity; reading effects by educational purpose renders those allocations empirically visible \cite{knox2020machine,perrotta2020deep}.

Applied here, the framework foregrounds three concerns: which educational purposes show gains, where evidence remains thin, and how the observed distribution bears on agency and participation. We estimate purpose-level effects of student-facing LLM interventions (RQ2) and examine how intervention characteristics are associated with effects \emph{within each purpose} (RQ3).

\section{Methodology}
\label{sec_methodology}

\begin{figure}[t]
\centering
\includegraphics[width=0.92\textwidth]{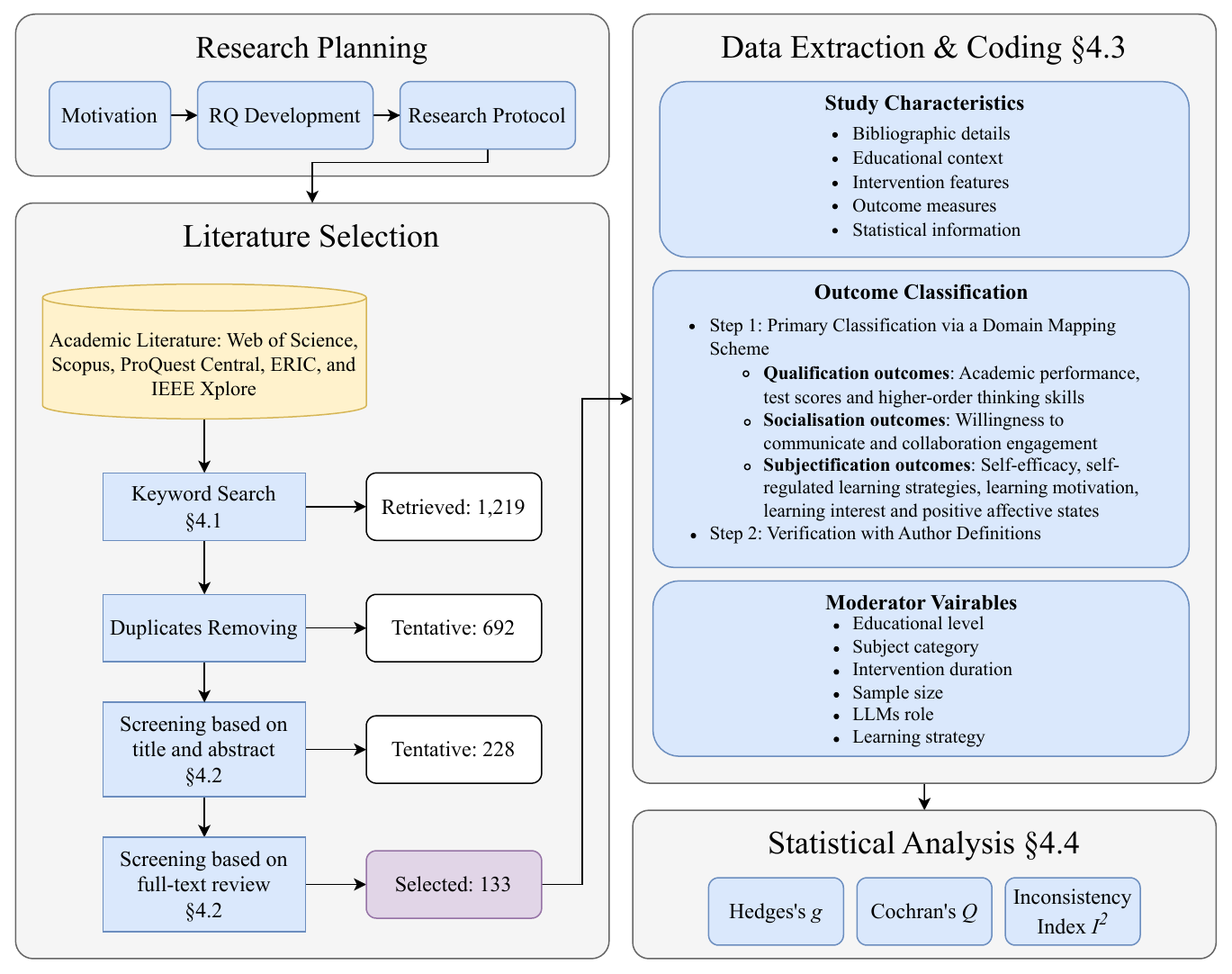}
\caption{An overview of our meta-analysis framework. 
}
\label{fig:prisma}
\end{figure}

To address these questions, we conducted a systematic meta-analysis following PRISMA 2020 guidelines (Figure~\ref{fig:prisma}).

For RQ1, we extracted descriptive characteristics of all eligible studies to map the current landscape of LLM interventions in education. For RQ2, we classified outcomes into Biesta's three domains, qualification, socialisation, and subjectification, and we estimated the overall effects within each. For RQ3, we coded intervention features (educational level, subject category, duration, sample size, LLM role and learning strategy) as moderators and tested their influence on effect sizes.
The methodology section proceeds as follows: \S\ref{subsec_search_strategy} outlines the search strategy, \S\ref{subsec_inclusion_and_exclusion_criteria} specifies inclusion and exclusion criteria, \S\ref{subsec_data_extraction_and_coding} details data extraction and coding (including the operationalisation of Biesta's framework), and \S\ref{subsec_statistical_analysis} describes effect size computation and statistical models.

\subsection{Search Strategy}
\label{subsec_search_strategy}

We conducted a systematic literature search, which was completed on March 18, 2025, across five major academic databases: Web of Science, Scopus, ProQuest Central, ERIC, and IEEE Xplore. The first four databases were selected because they are widely used in educational research and frequently appear in systematic reviews~\cite{heung2025chatgpt, deng2025does, liu2025impact}, while IEEE Xplore was included to capture studies from computer science, where many large language model (LLM) applications are published.

The search covered publications from November 2022, when ChatGPT was first released, through March 2025. We used the following query:  
\textit{(``Generative AI'' OR ``Large Language Model'' OR ``LLM'' OR ``ChatGPT'' OR ``GPT-4'' OR ``GPT-3'' OR ``GPT-3.5'') AND (``Education'' OR ``Learning'' OR ``Teaching'') AND (``Experimental Group'' OR ``Control Group'' OR ``ChatGPT Group'' OR ``Randomised Controlled Trial'' OR ``RCT'' OR ``Quasi-experimental'' OR ``Mixed-methods'')}.

The initial search identified 1,219 records. After removing 527 duplicates, 692 articles were screened by title and abstract, with 464 excluded for being irrelevant to LLM learning outcomes or for lacking an experimental design. The remaining 228 articles underwent full-text review. Of these, 95 were excluded for reasons such as non-experimental design, non-student participants, absence of a control group, inconsistent learning conditions, or insufficient quantitative data. The final dataset comprised 133 studies, which were retained for meta-analysis.

\subsection{Inclusion and Exclusion Criteria}
\label{subsec_inclusion_and_exclusion_criteria}

To establish a consistent evidence base, we defined a set of inclusion and exclusion criteria adapted from prior meta-analyses in educational technology~\cite{wang2025effect, deng2025does, heung2025chatgpt} and refined for the specific context of LLM interventions. Two researchers independently applied these criteria during both abstract screening and full-text review, with disagreements resolved through discussion to minimise bias.
Studies were included if they met four conditions: (1) they were peer-reviewed publications; (2) written in English; (3) they examined LLM interventions specifically targeting student learning outcomes; and (4) they employed an experimental or quasi-experimental design that incorporated a control or comparison group.
Conversely, we excluded studies that failed to provide a valid basis for effect size estimation. This applied to (1) non-experimental or purely descriptive studies, (2) research involving non-student participants, (3) cases where control groups also used LLMs or differed in learning methods beyond LLM usage, and (4) studies lacking sufficient quantitative data (\eg missing sample sizes, means, standard deviations, or test statistics). These criteria ensured that the final dataset consisted of robust and comparable evidence suitable for meta-analytic synthesis of LLMs' effects on student learning.

\subsection{Data Extraction and Coding}
\label{subsec_data_extraction_and_coding}

Following eligibility screening, all included studies were systematically coded using a structured codebook to enable effect size computation and moderator analysis. Two researchers coded independently, and discrepancies were resolved through discussion until consensus was reached. 

\textbf{Study characteristics.} For each study we recorded bibliographic details (author, year, outlet, and region), educational context (educational level, subject domain, sample size, and intervention duration), intervention features (LLM role, LLM type and learning strategy), outcome measures, and statistical information required to compute effect sizes (group size, means, standard deviations, or reported test statistics). Independent comparisons between an experimental and a control group were treated as separate entries. Where multiple independent experiments were reported within the same study, each was coded separately. 

\textbf{Outcome classification.} To address RQ2, we classified learning outcomes from the included studies into Biesta's three domains: qualification, socialisation, and subjectification. Recognizing that these domains are conceptually interdependent and often overlap in practice~\cite{coelho2025understanding}, we developed a systematic, two-step coding procedure to ensure consistency and mitigate misclassification. The process was conducted by two researchers, with ambiguous cases reviewed jointly until consensus was reached.

Step 1: Primary Classification via a Domain Mapping Scheme. We began by developing a primary mapping scheme based on Biesta's core definitions. This scheme guided our initial classification of the outcomes reported in each study:
\begin{itemize}
    \item Qualification outcomes were defined as those concerning the acquisition of knowledge, skills, and competencies. This category primarily included measures of academic performance, test scores and higher-order thinking skills (\eg critical thinking, creativity, problem-solving ability).
    \item Socialisation outcomes were those related to the induction of students into social and cultural practices. This included measures of willingness to communicate and collaboration engagement (\eg participation in group discussions).
    \item Subjectification outcomes were those focused on the cultivation of learner autonomy, agency, and identity. This category included measures of self-efficacy, self-regulated learning strategies, learning motivation, learning interest, and positive affective states (\eg greater learning enjoyment).
\end{itemize}

Step 2: Verification with Author Definitions. Second, we verified our initial coding against the definitions provided by the original studies' authors. We prioritised how the authors themselves conceptualized their outcome variables to ensure our coding was faithful to the original research context. For example, if a study measured ``engagement'' using a validated scale that focused on students' sense of belonging and peer interaction, we coded it as Socialisation, whereas if it was measured by on-task time during a problem-solving exercise, it was coded as Qualification. This two-step process ensured that our classification was both systematically grounded in our theoretical framework and empirically faithful to the context of the original studies.

\textbf{Moderator variables.} To address RQ3, we coded intervention-level moderators, drawing on prior coding schemes used in recent meta-analyses~\cite{zheng2023effectiveness, wang2025effect}: educational level (higher, secondary, primary, kindergarten), subject domain (arts and music, computer science, economics, education and psychology, language and writing, mathematics, medical and health, science), intervention duration ($<$1 week, 1--4 weeks, 4--8 weeks, $>$8 weeks), sample size (1--50, 51--100, $>$100), LLM role (tutor, partner, learning tool), and learning strategy (personalised, project-based, problem-based, contextual, reflective, gamified). Coding decisions followed author descriptions of interventions; when strategies overlapped, we coded the dominant activity as defined by the study's primary research objectives or instructional emphasis.

\textbf{Independence of effect sizes.} To ensure statistical independence, we followed the approach used in prior meta-analyses~\cite{wang2025effect}:: (1) each experimental vs. control group comparison yielded one effect size; (2) multiple independent experiments within the same study were coded separately; and (3) when multiple sub-dimensions of the same construct were reported, effect sizes were aggregated into a single value for that domain. When studies reported outcomes across multiple educational purposes, separate effect sizes were extracted for each. 

\textbf{Missing and ambiguous information.} When critical variables such as intervention duration or sample size were not explicitly reported, we inferred values where possible from descriptions of sessions or cohorts; otherwise, values were coded as missing and excluded from moderator analyses. Subject domains were coded according to the content of the measured learning outcome rather than course titles, to reduce misclassification in interdisciplinary contexts.

\textbf{Reliability.} Coding reliability was assessed after independent coding. Cohen's $\kappa$ was 0.93, indicating excellent agreement. Remaining discrepancies were resolved with the involvement of a third researcher. Ultimately, the coding process yielded 188 independent effect sizes extracted from 133 studies.

\begin{table}[t]

\centering
\caption{Coding scheme of the moderator variables.}
\label{tab:moderators}
\resizebox{\textwidth}{!}{%
\begin{tabular}{p{3.3cm}|p{12cm}}
\hline
\textbf{Moderator} & \textbf{Description} \\ \hline
\textbf{Educational Level} &  1. Higher Education (ages over 18) \\
  & 2. Secondary Education (ages 14–17) \\
  & 3. Primary Education (ages 7–13) \\
  & 4. Kindergarten (ages below 6) \\ \hline
\textbf{Subject Category} 
  & 1. Arts and Music (includes creative arts and music education) \\
  & 2. Computer Science (includes programming, robotics, and related digital technologies) \\
  & 3. Economics (includes economics, business ethics and entrepreneurship) \\
  & 4. Education and Psychology (includes general education, instructional design, psychology, and educational technology) \\
  & 5. Language and Writing (includes academic writing, second language learning, debate, and writing pedagogy) \\
  & 6. Mathematics (includes mathematics and statistics) \\
  & 7. Medical and Health (includes medical education, nursing, pharmacy, and healthcare training) \\
  & 8. Science (includes chemistry, physics, science and sustainability) \\ \hline
\textbf{Intervention Duration} & 1. $<$ 1 week \\
  & 2. 1–4 weeks \\
  & 3. 4–8 weeks \\
  & 4. $>$ 8 weeks \\ \hline
\textbf{Sample Size} &  1. 1–50 \\
  & 2. 51–100 \\
  & 3. $>$ 100 \\ \hline
\textbf{LLM Role} &  
1. Intelligent tutor (serves mainly as an instructor, offering tailored guidance, evaluative feedback, and structured learning support)~\cite{lin2023artificial} \\
& 2. Intelligent partner (acts as a peer-like collaborator, engaging in joint discussions and cooperative tasks to foster learning)~\cite{li2024expert} \\
& 3. Intelligent learning tool (functions as a utility that supplies resources, information, or task aids without engaging in social interaction)~\cite{pan2023exploring} \\ \hline

\textbf{Learning Strategy} &  
1. personalised learning (tailors instruction and resources to match each learner's needs, preferences, and pace)~\cite{escalante2023ai} \\
& 2. Project-based learning (centres on long-term tasks where students investigate and create products around complex questions)~\cite{perifanou2025collaborative} \\
& 3. Problem-based learning (builds understanding through tackling authentic, real-world challenges)~\cite{hamid2023exploratory} \\
& 4. Contextual learning (links new knowledge to specific, authentic situations to enhance relevance)~\cite{fu2023improving} \\
& 5. Reflective learning (encourages learners to think critically about their progress and make informed improvements)~\cite{samuel2025enhancing} \\
& 6. Gamified approach (applies game-like elements such as points or rewards to increase engagement)~\cite{kassenkhan2025gamification} \\ \hline

\end{tabular}
}
\end{table}

\subsection{Statistical Analysis}
\label{subsec_statistical_analysis}

All statistical analyses were conducted in R using the metafor package~\cite{viechtbauer2010conducting}. Most included studies reported post-test means and standard deviations for experimental and control groups, which allowed us to calculate standardized mean difference effect sizes. To correct for the small-sample bias inherent in Cohen's $d$, we used Hedges' $g$ ~\cite{hedges1981distribution}, defined as:
\begin{equation}\label{eq:cohen_d_block}
    d = \left(1 - \frac{3}{4N - 9}\right)  \frac{\bar{X}_1 - \bar{X}_2}{SD_{\text{pooled}}} ,
\end{equation} 
where $\bar{X}_1$ and $\bar{X}_2$ are the group means, $SD_{\text{pooled}}$ is the pooled standard deviation, and $N$ is the total sample size. When means or standard deviations were not reported, equivalent statistics (\eg $t$, $F$, or $\chi^2$) were converted into effect sizes using the formulas of \citet{lipsey2001practical}. To facilitate interpretation, the calculated effect sizes were interpreted following the extended guidelines proposed by~\cite{sawilowsky2009new}: very small (0.1), small (0.2), medium (0.5), large (0.8), very large (1.2), and huge (2.0). Heterogeneity was assessed using Cochran's $Q$ and the inconsistency index $I^2$. A $Q$-test $p$-value below 0.10 was taken as evidence of significant heterogeneity, in which case a random-effects model was applied for the meta-analysis~\cite{zhang2024effects}. The $I^2$ statistic further quantified the degree of heterogeneity, with 0–25\% interpreted as low, 25–75\% as moderate, and 75–100\% as high heterogeneity \citep{higgins2003measuring}.

\section{Results}
\label{sec_results}

\subsection{Descriptive Statistics: A Rapidly Growing but Skewed Research Landscape}

\begin{table}[t] 
\centering \caption{Descriptive statistics of included studies ($n = 133$).} 
\label{tab:source_publications} \resizebox{\textwidth}{!}{%
\begin{tabular*}{\textwidth}{@{\extracolsep{\fill}}llcc}
\toprule \textbf{Category} & \textbf{Subcategory} & \textbf{Frequency} & \textbf{Percent} \\ \midrule Year of publication & 2024 & 72 & 54.13 \\ & 2025 & 51 & 38.35 \\ & 2023 & 10 & 7.52 \\ \midrule Type of publication & Journal article & 116 & 87.22 \\ & Conference paper & 17 & 12.78 \\ \midrule Study location & Asia & 104 & 78.2 \\ & Europe & 14 & 10.53 \\ & North America & 7 & 5.36 \\ & South America & 6 & 4.51 \\ & Africa & 2 & 1.50 \\ \midrule Educational Level & Higher Education & 108 & 81.20 \\ & Secondary Education & 16 & 12.03 \\ & Primary Education & 8 & 6.02 \\ & Kindergarten & 1 & 0.75 \\ \midrule Subject & Language and Writing & 49 & 36.84 \\ & Computer Science & 32 & 24.06 \\ & Medical and Health & 16 & 12.03 \\ & Science & 13 & 9.77 \\ & Education and Psychology & 11 & 8.27 \\ & Mathematics & 5 & 3.76 \\ & Economics & 5 & 3.76 \\ & Arts and Music & 2 & 1.50 \\ \midrule LLM Model & ChatGPT & 116 & 87.22 \\ & Other LLM & 17 & 12.78 \\ \midrule Intervention Duration & $>$ 8 weeks & 42 & 31.58 \\ & 4--8 weeks & 40 & 30.08 \\  & 1--4 weeks & 28 & 21.05 \\ & $<$ 1 week & 23 & 17.29 \\ \midrule Sample Size & 51--100 & 68 & 51.13 \\ & 1--50 & 39 & 29.32 \\ & $>$100 & 26 & 19.55 \\ \midrule LLM Role & Intelligent tutor & 60 & 45.11 \\ & Intelligent learning tool & 44 & 33.08 \\ & Intelligent partner & 29 & 21.80 \\ \midrule Learning Strategy & personalised learning & 39 & 29.32 \\ & Problem-based learning & 26 & 19.55 \\ & Reflective learning & 25 & 18.80 \\ & Contextual learning & 21 & 15.79 \\ & Project-based learning & 17 & 12.78 \\ & Gamified approach & 5 & 3.76 \\ \bottomrule \end{tabular*} } \end{table}

The descriptive profile of the 133 included studies, as summarized in Table~\ref{tab:source_publications}, points to a research field that is both expanding at remarkable speed and unevenly distributed. 
The publication timeline illustrates its novelty and acceleration: more than half of all studies appeared in 2024 (54.13\%), with a further 38.35\% published in the first quarter of 2025, a surge that closely followed the public release of ChatGPT in late 2022.

This growth, however, is accompanied by pronounced structural imbalances. 

\textbf{Geographical concentration.} The evidence base is heavily skewed towards Asia, which accounts for 78.20\% of all studies. By contrast, North America (5.36\%) and Europe (10.53\%) are strikingly under-represented. As a result, the current literature reflects predominantly Asian educational settings and cultural assumptions, raising questions about the transferability of findings to other contexts. 

\textbf{Educational level.} Research is also disproportionately concentrated in higher education (81.20\%). Far fewer studies investigate secondary (12.03\%) or primary (6.02\%) education, levels where learners' developmental needs and classroom ecologies differ substantially. This leaves a major evidence gap regarding how LLMs function in earlier stages of schooling. 

\textbf{Disciplinary focus.} Studies cluster around domains where the relevance of LLMs is most immediate. Language and writing (36.84\%) and computer science (24.06\%) dominate, whereas mathematics (3.76\%), arts and music (1.50\%) are rarely addressed. Such patterns suggest that the field is still largely exploring “low-hanging fruit” applications, with much less known about how LLMs might support learning outside language-centric or computational tasks. 

\textbf{Intervention design.} Encouragingly, the field is beginning to move beyond short, one-off trials. A majority of interventions extend over four weeks or longer (61.66\%), and nearly half (45.11\%) position LLMs as an intelligent tutor embedded within broader pedagogical models such as personalised learning (29.32\%) or problem-based learning (19.55\%). This reflects a shift towards more authentic, sustained integrations of LLMs into teaching practice. 

\textbf{Model type.} However, the evidence base remains overwhelmingly dominated by a single model. Nearly nine in ten studies (87.22\%) relied exclusively on ChatGPT, while only 12.78\% involved other large language models, such as Gemini, LLaMA, ERNIE Bot, and Qwen. This heavy reliance limits the generalisability of findings and underscores the need for comparative studies across different LLM architectures. 

In summary, the literature on LLMs in education is growing at an exceptional pace but remains defined by university-based research in Asia, concentrated in language and computer science. These descriptive patterns provide critical context for the effect-size estimates reported in the next section and highlight limitations that will be further interrogated in the discussion. 

\begin{table}[t]
\centering
\caption{Results of overall effects analysis of LLM on students' learning outcomes.}
\label{tab:overall_effects}

\resizebox{\textwidth}{!}{%
\begin{tabular}{lccccccccc}
\toprule
\textbf{Learning outcome} & \textbf{\textit{k}} & \textbf{\textit{g}} & \multicolumn{2}{c}{\textbf{95\% CI}} & \multicolumn{2}{c}{\textbf{Two-tailed test}} & \multicolumn{3}{c}{\textbf{Test of homogeneity}} \\
\cmidrule(lr){4-5}\cmidrule(lr){6-7}\cmidrule(lr){8-10}
 &  &  & \textit{LL} & \textit{UL} & \textit{Z} & \textit{P} & \textit{Q} & \textit{df(Q)} & \textit{I\textsuperscript{2}} \textit{p} \\
\midrule
Qualification   & 123 & 0.751 & 0.635 & 0.867 & 12.7 & $<.0001$ & 1008.69 & 122 & 88.23\% $<.0001$ \\
Socialisation             & 11  & 0.745 & 0.429 & 1.06 & 4.63 & $<.0001$ & 53.823 & 10  & 80.82\% $<.0001$ \\
Subjectification  & 54  & 0.654 & 0.506 & 0.802 & 8.66 & $<.0001$ & 274.30 & 53  & 85.57\% $<.0001$ \\
\bottomrule
\end{tabular}%
}

\vspace{0.3em}

\small \textit{Note.} $k$ = number of independent studies; $g$ = mean effect size; CI = confidence interval.
\end{table}

\subsection{Overall Effects: LLMs Enhance Learning, but Unevenly Across Domains}

To answer RQ2, we estimated the overall effects of LLM-based interventions across Biesta's three educational domains (Table~\ref{tab:overall_effects}). Across all domains, results indicate significant positive impacts, though the magnitude and stability of these effects differ. Given the high degree of heterogeneity across studies ($I^{2} = 80.8\%\text{--}88.2\%, p < .0001$), random-effects models were applied throughout.

\begin{itemize}
    \item \textit{Qualification}. The strongest and most consistent effects emerged for qualification, encompassing knowledge acquisition, skill development, and academic performance. The estimated effect size was large and robust ($g = 0.751$, 95\% CI [0.635, 0.867], $p < .0001$), underscoring LLMs' clear value in supporting measurable academic achievement. 
    \item \textit{Socialisation}. LLMs also exerted a significant positive influence on socialisation outcomes ($g = 0.745$, 95\% CI [0.429, 1.060], $p < .0001$), which include communication, collaboration, and participation in learning communities. However, the relatively wide confidence interval points to greater variability across studies, suggesting that socialisation gains are more context-dependent. 
    \item \textit{Subjectification}. The effect on subjectification, defined here as the development of learner autonomy, agency, and identity, was positive but smaller ($g = 0.654$, 95\% CI [0.506, 0.802], $p < .0001$). While significant, this finding highlights a more fragile and uneven evidence base for LLMs' ability to foster students' personal becoming. 
\end{itemize}

Taken together, these findings confirm that LLMs are effective in enhancing learning but do so unevenly. Their most reliable contributions lie in boosting qualification, whereas their effects on socialisation and subjectification, though positive, are less consistent and less pronounced. This hierarchy of impacts underscores the need for our subsequent moderator analyses, which examine the conditions under which LLMs can more effectively support the broader social and developmental purposes of education. 

\subsection{Moderator Analyses: When and How LLMs Work}

\begin{table}[t]
\centering
\caption{Moderator analysis for Qualification.}
\label{tab:Qualification moderator}
\begin{tabular*}{\textwidth}{@{\extracolsep{\fill}}lcccccc}
\toprule
\textbf{Moderator} & $k$ & $g$ (95\% CI) & $z$ & df($Q$) & $Q_B$ & $p$-value \\
\midrule

\textbf{Educational Level} & & & & 3 & 0.436 & 0.933 \\
\quad 1. Higher Education & 98 & 0.771 [0.639, 0.903] & 11.45*** & & & \\
\quad 2. Secondary Education & 16 & 0.691 [0.376, 1.007] & 4.29*** & & & \\
\quad 3. Primary Education & 8 & 0.651 [0.194, 1.108] & 2.79** & & & \\
\quad 4. Kindergarten & 1 & 0.651 [-0.643, 1.945] & 0.99 & & & \\

\textbf{Subject Category} & & & & 7 & 10.859 & 0.145 \\
\quad 1. Arts \& Music & 2 & 1.147 [0.278, 2.017] & 2.59** & & & \\
\quad 2. Computer Science & 32 & 0.638 [0.415, 0.861] & 5.60*** & & & \\
\quad 3. Economics & 4 & 1.368 [0.713, 2.024] & 4.09*** & & & \\
\quad 4. Education \& Psychology & 10 & 0.782 [0.352, 1.211] & 3.56*** & & & \\
\quad 5. Language & 41 & 0.679 [0.483, 0.875] & 6.79*** & & & \\
\quad 6. Mathematics & 5 & 0.342 [-0.236, 0.919] & 1.16 & & & \\
\quad 7. Medical \& Health & 16 & 0.812 [0.495, 1.128] & 5.03*** & & & \\
\quad 8. Science & 13 & 1.058 [0.714, 1.402] & 6.02*** & & & \\

\textbf{Intervention Duration} & & & & 3 & 18.091 & 0.0004*** \\
\quad 1. < 1 week & 21 & 0.622 [0.355, 0.889] & 4.56*** & & & \\
\quad 2. 1–4 weeks & 27 & 0.389 [0.158, 0.620] & 3.30*** & & & \\
\quad 3. 4–8 weeks & 38 & 0.805 [0.640, 0.970] & 7.86*** & & & \\
\quad 4. > 8 weeks & 37 & 1.024 [0.827, 1.221] & 10.21*** & & & \\

\textbf{Sample Size} & & & & 2 & 1.017 & 0.602 \\
\quad 1. 1–50 & 37 & 0.787 [0.560, 1.013] & 6.81*** & & & \\
\quad 2. 51–100 & 62 & 0.781 [0.619, 0.943] & 9.44*** & & & \\
\quad 3. > 100 & 24 & 0.639 [0.391, 0.886] & 5.06*** & & & \\

\textbf{LLM Role} & & & & 2 & 13.341 & 0.0013** \\
\quad 1. Intelligent learning tool & 37 & 0.436 [0.234, 0.638] & 4.24*** & & & \\
\quad 2. Intelligent partner & 29 & 0.842 [0.611, 1.074] & 7.12*** & & & \\
\quad 3. Intelligent tutor & 57 & 0.902 [0.741, 1.062] & 11.01*** & & & \\

\textbf{Learning Strategy} & & & & 5 & 6.011 & 0.305 \\
\quad 1. Contextual learning & 21 & 0.990 [0.708, 1.271] & 6.90*** & & & \\
\quad 2. Gamified approach & 5 & 0.654 [0.091, 1.217] & 2.28* & & & \\
\quad 3. personalised learning & 34 & 0.742 [0.523, 0.961] & 6.64*** & & & \\
\quad 4. Problem-based learning & 26 & 0.568 [0.314, 0.822] & 4.38*** & & & \\
\quad 5. Project-based learning & 16 & 0.901 [0.571, 1.231] & 5.35*** & & & \\
\quad 6. Reflective learning & 21 & 0.669 [0.393, 0.945] & 4.76*** & & & \\

\bottomrule
\end{tabular*}

\small\textit{Note.} $z$ = z-value for $g$; 
$Q_B$ = between-group heterogeneity. \textit{p-value:} * $p<.05$, ** $p<.01$, *** $p<.001$.
\end{table}

To address RQ3, we examined moderators that shape the impact of LLM interventions. Across domains, the analyses showed that how an LLM is used matters as much as whether it is used at all. Two factors consistently emerged as decisive: the pedagogical role assigned to the LLM and the duration of the intervention. Programs that positioned LLMs as active tutors, and that sustained their use over eight weeks or longer, achieved the largest and most reliable gains. Other moderators, including subject domain, pedagogical strategy, and sample size exerted more domain-specific effects.

\subsubsection{Qualification: Strongest Gains with Tutor Roles and Sustained Use}

For qualification outcomes (see Table~\ref{tab:Qualification moderator}), both the LLM's role ($Q_{B} = 13.341, p = .0013$) and intervention duration ($Q_{B} = 18.091, p < .001$) were highly significant moderators. The hierarchy of effects was clear: LLMs framed as interactive tutors ($g = 0.902$) or collaborative partners ($g = 0.842$) generated substantially larger improvements in performance than when they functioned as passive tools ($g = 0.436$). Duration also mattered in a ``dose–response'' fashion: interventions longer than eight weeks produced the largest effects ($g = 1.024$), while those lasting only 1–4 weeks were far weaker ($g = 0.389$).

Notably, qualification effects were otherwise stable: no significant moderation was found by education level, subject area, sample size, or learning strategy. This suggests that long-term, tutor-style deployments of LLMs are broadly effective for enhancing knowledge and skills across diverse contexts. In summary, for boosting knowledge and skills, LLMs are most potent when they are designed to act as persistent, active pedagogical agents rather than as one-off, functional utilities.

\begin{table}[t]
\centering
\caption{Moderator analysis for Socialisation.}
\label{tab:Socialization moderator}
\begin{tabular*}{\textwidth}{@{\extracolsep{\fill}}lcccccc}
\toprule
\textbf{Moderator} & $k$ & $g$ (95\% CI) & $z$ & df($Q$) & $Q_B$ & $p$-value \\
\midrule
\textbf{Education Level} & & & & 3 & 6.40 & 0.0939 \\
\quad 1. Higher Education & 8 & 0.860 [0.538, 1.181] & 5.25*** & & & \\
\quad 2. Secondary Education & 1 & 0.501 [-0.367, 1.369] & 1.13 & & & \\
\quad 3. Primary Education & 1 & 1.124 [0.187, 2.061] & 2.35* & & & \\
\quad 4. Kindergarten & 1 & -0.299 [-2.131, -0.187] & -2.34* & & & \\

\textbf{Subject Category} & & & & 1 & 1.447 & 0.229 \\
\quad 1. Language & 8 & 0.861 [0.499, 1.223] & 4.66*** & & & \\
\quad 2. Science  & 3 & 0.436 [-0.153, 1.026] & 1.45 & & & \\

\textbf{Intervention Duration} & & & & 3 & 16.470 & 0.0009** \\
\quad 1. < 1 week & 1 & 0.088 [-0.590, 0.766] & 0.25 & & & \\
\quad 2. 1–4 weeks & 2 & 0.152 [-0.323, 0.626] & 0.63 & & & \\
\quad 3. 4–8 weeks & 4 & 0.767 [0.432, 1.102] & 4.48*** & & & \\
\quad 4. > 8 weeks & 4 & 1.204 [0.860, 1.547] & 6.87*** & & & \\

\textbf{Sample Size} & & & & 2 & 3.818 & 0.1482 \\
\quad 1. 1–50 & 2 & 1.207 [0.461, 1.953] & 3.17** & & & \\
\quad 2. 51–100 & 8 & 0.571 [0.234, 0.908] & 3.32*** & & & \\
\quad 3. > 100 & 1 & 1.245 [0.392, 2.098] & 2.86** & & & \\

\textbf{LLM Role} & & & & 2 & 1.125 & 0.5699 \\
\quad 1. Intelligent learning tool & 3 & 0.998 [0.383, 1.613] & 3.18** & & & \\
\quad 2. Intelligent partner & 5 & 0.714 [0.233, 1.196] & 2.91** & & & \\
\quad 3. Intelligent tutor & 3 & 0.523 [-0.119, 1.166] & 1.60 & & & \\

\textbf{Learning Strategy} & & & & 3 & 12.752 & 0.0052** \\
\quad 1. Contextual learning & 3 & 0.168 [-0.260, 0.597] & 0.77 & & & \\
\quad 2. personalised learning & 2 & 1.299 [0.824, 1.775] & 5.35*** & & & \\
\quad 3. Project-based learning & 3 & 0.718 [0.295, 1.142] & 3.33*** & & & \\
\quad 4. Reflective learning & 3 & 0.921 [0.457, 1.384] & 3.89*** & & & \\

\bottomrule
\end{tabular*}
\vspace{0.3em}

\small\textit{Note.} $z$ = z-value for $g$; 
$Q_B$ = between-group heterogeneity. \textit{p-value:} * $p<.05$, ** $p<.01$, *** $p<.001$.
\end{table}

\subsubsection{Socialisation: Efficacy Hinges on Sustained Use and Structured Pedagogy}

For socialisation, which involves skills like communication and collaboration, the analysis revealed that intervention duration ($Q_{B}=16.470$, $p=.0009$) and learning strategy ($Q_{B}=12.752$, $p=.0052$) were the decisive moderators (see Table~\ref{tab:Socialization moderator}). A clear ``dose--response'' relationship emerged for duration: interventions lasting more than eight weeks yielded exceptionally large gains ($g=1.204$), whereas those under four weeks produced negligible effects. Similarly, pedagogical structure was critical; actively engaging approaches like personalised learning ($g=1.299$) and reflective learning ($g=0.921$) generated strong outcomes, in sharp contrast to the ineffectiveness of contextual learning ($g=0.168$).

Notably, several other factors were not significant. The LLM's role, subject category, and sample size did not moderate the effects on socialisation. While a marginal trend was observed for education level ($p=.094$), the sparse data in categories outside of Higher Education prevent any firm conclusions. In summary, fostering socialisation with LLMs depends less on the tool itself and more on its integration into long-term, structurally-sound learning activities.

\begin{table}[t]
\centering
\caption{Moderator analysis for Subjectification.}
\label{tab:Subjectification moderator}
\begin{tabular*}{\textwidth}{@{\extracolsep{\fill}}lcccccc}
\toprule
\textbf{Moderator} & $k$ & $g$ (95\% CI) & $z$ & df($Q$) & $Q_B$ & $p$-value \\
\midrule
\textbf{Education Level} & & & & 2 & 3.178 & 0.204 \\
\quad 1. Higher Education & 44 & 0.716 [0.552, 0.880] & 8.57*** & & & \\
\quad 2. Secondary Education & 7 & 0.339 [-0.047, 0.725] & 1.72 & & & \\
\quad 3. Primary Education & 3 & 0.570 [-0.052, 1.192] & 1.80 & & & \\

\textbf{Subject Category} & & & & 7 & 2.042 & 0.958 \\
\quad 1. Arts \& Music & 1 & 0.305 [-0.773, 1.383] & 0.55 & & & \\
\quad 2. Computer Science & 14 & 0.551 [0.242, 0.861] & 3.49*** & & & \\
\quad 3. Economics & 2 & 0.624 [-0.201, 1.448] & 1.48 & & & \\
\quad 4. Education \& Psychology & 5 & 0.661 [0.147, 1.175] & 2.52* & & & \\
\quad 5. Language & 23 & 0.731 [0.489, 0.972] & 5.93*** & & & \\
\quad 6. Mathematics & 1 & 0.934 [-0.241, 2.109] & 1.56 & & & \\
\quad 7. Medical \& Health & 5 & 0.781 [0.266, 1.296] & 2.97** & & & \\
\quad 8. Science & 3 & 0.451 [-0.201, 1.103] & 1.36 & & & \\

\textbf{Intervention Duration}& & & & 3 & 12.283 & 0.0065** \\
\quad 1. $<$ 1 week & 10 & 0.542 [0.223, 0.861] & 3.33*** & & & \\
\quad 2. 1–4 weeks & 12 & 0.291 [0.001, 0.581] & 1.96* & & & \\
\quad 3. 4–8 weeks & 15 & 0.668 [0.407, 0.928] & 5.02*** & & & \\
\quad 4. $>$ 8 weeks & 17 & 0.955 [0.710, 1.200] & 7.63*** & & & \\

\textbf{Sample Size} & & & & 2 & 9.019 & 0.011* \\
\quad 1. 1–50 & 14 & 1.044 [0.745, 1.342] & 6.85*** & & & \\
\quad 2. 51–100 & 26 & 0.590 [0.391, 0.789] & 5.81*** & & & \\
\quad 3. $>$100 & 14 & 0.470 [0.220, 0.720] & 3.68*** & & & \\

\textbf{LLM Role} & & & & 2 & 6.062 & 0.048* \\
\quad 1. Intelligent learning tool & 19 & 0.503 [0.264, 0.743] & 4.12*** & & & \\
\quad 2. Intelligent partner & 11 & 0.476 [0.156, 0.796] & 2.92** & & & \\
\quad 3. Intelligent tutor & 24 & 0.857 [0.640, 1.074] & 7.73*** & & & \\

\textbf{Learning Strategy} & & & & 5 & 4.346 & 0.501 \\
\quad 1. Contextual learning & 9 & 0.759 [0.383, 1.135] & 3.95*** & & & \\
\quad 2. Gamified approach & 1 & 0.112 [-0.936, 1.160] & 0.21 & & & \\
\quad 3. personalised learning & 13 & 0.676 [0.374, 0.979] & 4.38*** & & & \\
\quad 4. Problem-based learning & 8 & 0.384 [-0.008, 0.775] & 1.92 & & & \\
\quad 5. Project-based learning & 7 & 0.566 [0.154, 0.978] & 2.69** & & & \\
\quad 6. Reflective learning & 16 & 0.796 [0.517, 1.074] & 5.61*** & & & \\
\bottomrule
\end{tabular*}
\vspace{0.3em}

\small\textit{Note.} $z$ = z-value for $g$; 
$Q_B$ = between-group heterogeneity. \textit{p-value:} * $p<.05$, ** $p<.01$, *** $p<.001$.

\end{table}

\subsubsection{Subjectification: Tutor Roles and Small-Scale, Long-Term Programs}

For subjectification, the development of learner agency and autonomy, both the role of the LLM ($Q_{B} = 6.062, p = .048$) and duration of intervention ($Q_{B} = 12.283, p = .0065$) again proved decisive (see table~\ref{tab:Subjectification moderator}). Framing the LLM as a tutor ($g = 0.857$) was significantly more effective than using it as a partner ($g = 0.476$) or tool ($g = 0.503$). Programs extending beyond eight weeks ($g = 0.955$) produced the strongest outcomes. 
Crucially, sample size moderated effects ($Q_{B} = 9.019, p = .011$). Studies with 1–50 participants reported very large impacts on learner agency ($g = 1.044$), while effects diminished sharply in cohorts larger than 100 ($g = 0.470$). By contrast, education level, subject category, and learning strategy were not significant moderators. This points to a structural constraint: fostering subjectification with LLMs appears most feasible in small, highly structured, long-term programs, while scalability remains a challenge. 

Together, these moderator analyses reveal a nuanced picture. For qualification, the story is clear: LLMs are most effective when used as sustained tutors across contexts. For socialisation, benefits hinge on alignment with language-rich domains, reflective pedagogies, and small-group settings. For subjectification, positive effects emerge under highly structured, tutor-like designs but are strongest in small-scale, long-term interventions. Across domains, the key lesson is that implementation design, not simply access, determines whether LLMs contribute to performance, participation, or autonomy.

\section{Discussion}
\label{sec:discussion}

\subsection{Contributions: Findings in Context}

This section situates our main findings relative to some closest meta-analyses (Table~\ref{tab:relatedwork}) and explains what is substantively novel. 

\emph{Purpose-level effects clarify what prior syntheses left implicit.}
Prior reviews more converge on “qualification” gains~\cite{deng2025does,sun2024does,liu2025effects,wang2025effect,liu2025impact,qu2025generative} and sometimes report engagement or affect~\cite{heung2025chatgpt,hu2025evaluating,zhu2025exploring,xia2025impact}, but they do not synthesise interactional participation and treat autonomy-related constructs only tangentially. By estimating effects at the level of qualification, socialisation, and subjectification, we find that all three purposes improve on average in our sample, with subjectification positive but less stable. This moves interpretation from ``LLMs raise scores'' toward ``which educational purposes benefit,'' a distinction not captured by Bloom-aligned groupings.

\emph{Within-purpose moderators help reconcile mixed results.}
 prior syntheses list design and context factors, but moderator tests typically pool mixed outcomes~\cite{hu2025evaluating, zhu2025exploring}, making it unclear whether conditions differentially support participation or autonomy. We test moderators within each purpose. At a high level, we estimate larger qualification effects for tutor-like, multi-week deployments; socialisation effects that depend critically on the intervention's duration and pedagogical structure; and subjectification gains that appear under tutor-like, longer-duration, smaller-scale programs. These patterns align with emerging experimental evidence underscoring the importance of pedagogical design in shaping LLMs' educational impact. For example, when students are given unguided access to a powerful LLM, it can function as a "crutch" that undermines long-term learning, leaving them worse off than peers without access~\cite{bastani2025generative}. By contrast, when an AI is deliberately designed as a tutor with pedagogical safeguards, it can outperform even well-established active learning techniques and enhance student engagement~\cite{kestin2025ai, bastani2025generative}. The surrounding study context also matters: combining LLM use with a traditional tactic like note-taking leads to greater comprehension and retention than LLM use alone, likely because note-taking elicits the deeper cognitive engagement essential for memory formation~\cite{kreijkes2025effects}.

\emph{Integrating seemingly divergent findings across adjacent reviews.}
Existing meta-analyses present apparently inconsistent conclusions: some emphasise substantial performance improvements in higher education~\cite{sun2024does, liu2025effects, liu2025impact}, others find limited effects on higher-order cognition~\cite{deng2025does, qu2025generative}, while others highlight behavioural or emotional engagement~\cite{hu2025evaluating, heung2025chatgpt}. Viewed through the lens of educational purposes, these findings are compatible: qualification effects are robust~\cite{scarlatos2025training, noy2023experimental, alneyadi2023chatgpt}; socialisation gains emerge from the pedagogical design, specifically, long-term and structured interaction, rather than the subject context~\cite{kovari2025systematic, hu2025enhancing, wei2025effects}; subjectification requires designs that cultivate agency rather than answer-seeking~\cite{dai2025students, kumar2023impact, chen2025more, alm2024exploring}. By aligning effect estimates with purpose and conducting moderator tests within each domain, our study integrates results that previous work examined in isolation and clarifies the conditions under which they hold.

\subsection{Descriptive Landscape and Coverage Gaps (RQ1)}
\label{sec_RQ1_discussion}

What makes these descriptives consequential is not only who and where is studied, but which \textit{design imaginaries} and \textit{measurement choices} become dominant as a result. First, an ecological concentration in Asian higher education and in language or computer science courses tends to favour designs that fit individual assignment workflows and course management platforms~\cite{belkina2025implementing, zawacki2019systematic, marzano2025generative}. This privileges tutor-style deployments and unit-length exposures, and it naturally leads instructors to select outcomes that are easy to capture at that grain size. Second, those outcomes are predominantly qualification indicators, such as tests and task accuracy. They are appropriate for mastery claims, but they underexpose interactional participation and agency, which are central for socialisation and subjectification~\cite{mouta2025agency, joseph2025rethinking, uanachain2025generative}. In other words, what is being measured follows from how the systems are being used and where they are used.

Two interpretive cautions follow from this structure. 

\emph{Transferability.} Because the corpus analysed in this study is anchored in Asian higher education and language or computer science, extrapolation to primary and secondary schooling or to domains like mathematics and arts remains uncertain. The social organisation of classrooms and the role of talk differ substantially in those settings, which can change how socialisation and subjectification show up when LLMs are present. 

\emph{Construct asymmetry.} When most deployments use tutor roles and assess near-term mastery, average effects will mainly reflect qualification~\cite{selwyn2016technology,williamson2022education}. This does not imply that LLMs cannot support participation or autonomy; rather, it shows that current studies rarely instrument for those constructs. For human–computer interaction, the implication is straightforward: unless interactional and agency measures are foregrounded as outcomes, the impact conveyed to designers and policymakers will remain unduly narrow.

These observations do not diminish the value of existing studies. They indicate what today's evidence is optimised to detect and where blind spots persist. In §\ref{sec_RQ2_discussion} and §\ref{sec_RQ3_discussion}, we therefore apply these interpretive filters, treating strong qualification estimates as securely grounded in the present design space, while viewing socialisation and subjectification estimates as dependent on the explicit design and measurement of interaction and agency.

\subsection{Purpose-Level Effects (Qualification, Socialisation, Subjectification) (RQ2)}
\label{sec_RQ2_discussion}

We estimate positive average effects within each purpose: qualification \(g=0.751\), socialisation \(g=0.745\), and subjectification \(g=0.654\).

\emph{Qualification.} Gains are higher on average for outcomes aligned with knowledge and skill acquisition. This pattern is consistent with current student-facing uses and assessments of LLMs, where item-level, exemplars, and stepwise hints align with mastery tests and rubric-based assessments~\cite{lee2024harnessing, yavuz2025utilizing, zhao2024generative}.

\emph{Socialisation.} Positive average effects indicate that LLM-supported activities can enhance participation and collaboration, but these gains appear more contingent on the intervention design. Measures often capture dialogue quality, contribution, or communicative clarity. Improvements are most pronounced when interaction is scaffolded, such as with reflective prompts or structured peer exchange, and these activities are sustained over longer periods~\cite{msambwa2025impact, zha2025colp, kovari2025systematic, wiboolyasarin2024synergizing}.

\emph{Subjectification.} Average gains in autonomy and agency are positive but less stable. Autonomy is not only a cognitive state; it is a relation to peers, tasks, norms, and tools. Designs that put learners in control, such as tutor-like systems that hold back the final answer and ask for justification, or workflows with plan–revise cycles and reflective journals, tend to show stronger signals of self-regulation and ownership~\cite{roe2024generative, dai2025students, mouta2025agency}. In contrast, tasks that reward only speed or correctness rarely track subjectification outcomes, and any benefits are typically superficial~\cite{bastani2025generative, kasneci2023chatgpt}. 

\emph{Heterogeneity and durability.} The positive averages sit atop substantial variability across studies. External trials are consistent with a common boundary condition: unguided answer use can inflate immediate scores yet depress delayed performance when AI is removed, whereas guardrailed tutoring and study tactics such as note-taking mitigate that risk \cite{bastani2025generative,kreijkes2025effects}. Distributional effects reported elsewhere suggest that lower-baseline readers may benefit more from summarisation tools, while higher-baseline readers can be harmed \cite{etkin2025differential}, and that over-reliance can weaken higher-order cognition \cite{zhai2024effects}. Viewed through Biesta's framework, these tensions are expected: privileging qualification without designing for participation and agency compromises socialisation and subjectification at follow-up.

\subsection{Moderator Effects Interpretation (RQ3)}
\label{sec_RQ3_discussion}

We modelled six moderators: three \emph{design} moderators (LLM role, intervention duration, learning strategy) and three \emph{context} moderators (educational level, subject category, sample size). Below we interpret patterns \emph{within} each purpose and draw out what they imply for interaction design and evaluation.

\emph{Qualification.} Effects were larger when deployments were tutor-like rather than answer-providing and when exposure was multi-week rather than one-shot. Two mechanisms plausibly underlie this pattern. First, \emph{allocation of initiative}: tutor-like designs request plans, rationales, and self-checks, which increase time-on-reasoning and reduce copy-through~\cite{van2023chatgpt, chi2014icap}. Second, \emph{feedback granularity}: stepwise hints and error-contingent prompts align with mastery rubrics, making gains legible to achievement measures~\cite{szymanski2025granular}. Learning strategy, level, subject, and sample size did not reliably differ in our model. A practical reading is that role and duration act as first-order levers for qualification, with other factors mostly bounding transfer rather than shifting the mean.

\emph{Socialisation.} The effects on socialisation were highly sensitive to the intervention's design. The key drivers were not the subject context but the pedagogical structure, particularly when interventions were sustained over time. The most effective designs provided strong \emph{interactional scaffolding} through strategies like reflective or personalised learning~\cite{lin2025enhancing, pratama2023revolutionizing}. This principle helps explain the high variability in outcomes: when tasks require perspective-taking, justification, or peer exchange, LLM support can enhance participation; however, in contexts with specialised norms (e.g., professional programs), generic conversational affordances may not align with expected practices, yielding mixed results~\cite{zhao2023beyond, ling2023domain}. Methodologically, a limitation of the current evidence is that many socialisation measures are coarse (e.g., counts of studies). Future work should pair designs that elicit dialogue with instruments that capture the \emph{quality}, not just the quantity, of participation.

\emph{Subjectification.} Gains in autonomy and agency appeared more often under tutor-like designs, extended duration, and small-scale programs; effects attenuate in large cohorts. Autonomy is not merely a cognitive state but a relation to peers, tasks, norms, and technology. It flourishes when students exercise disciplined choice, self-explanation, and responsibility within a structure~\cite{nopas2025algorithmic, grund2020facilitating}, consistent with Biesta's view that subjectification is freedom with limits, not from limits \cite{biesta2015beautiful}. The scalability challenge we observe is therefore principled: agency practices are harder to orchestrate at scale unless schools redesign roles, time, and assessment. 

\emph{Cross-cutting synthesis.} Across purposes, the same basic dimension recurs: \emph{who holds initiative when uncertainty arises}. Designs that keep the learner initiating, explaining, and revising tend to lift qualification and create conditions for socialisation and subjectification to register. Role and duration consistently shape outcomes, while strategy and subject matter only influence results in certain contexts. Level and sample size primarily constrain transfer and orchestration rather than shift effects.

\subsection{Subjectification in the Context of AI: Significance and Risks}

When access to explanations becomes almost effortless, we need to rethink what education is for. Our synthesis (RQ2) shows that outcomes linked to subjectification -- students' ability to take initiative and act with responsibility -- are generally positive but less stable than gains in knowledge or collaboration, and they depend more on how learning is designed and how much time is given. This pattern supports Biesta's view that becoming an active subject is not just about producing correct answers; it requires structured chances to choose, justify, and take ownership of decisions.

Our moderator analysis (RQ3) helps explain this vulnerability. Growth in agency tends to occur when LLMs are used in tutor-like ways over longer periods, and it weakens when designs focus on giving answers quickly or provide only brief exposure. The key mechanism is who holds the initiative: if the system supplies solutions too readily, students may lean on it, risking “automation bias” and losing confidence in their own judgement~\cite{darvishi2024impact, shukla2024ai}. When learners are encouraged to plan, explain, and revise, they practise agency and build resilience~\cite{smith2017encouraging, ahmadi2019study}. Classroom studies show the difference: unstructured answer use can boost immediate scores but undermine later performance once AI is removed, whereas note-taking and reflective prompts help sustain learning over time~\cite{kreijkes2025effects}.

The main risk is not the technology itself but how we position it in education. Systems designed only for fast answers can narrow teaching and learning to what is easiest to measure, sidelining opportunities for thoughtful judgement~\cite{selwyn2016technology, williamson2022education}. The implications are wide-ranging: philosophical (what kind of agency we want learners to have), pedagogical (how to design for growth as well as performance), and institutional (what schools choose to value as success).

\subsection{Publication Bias and Sensitivity Analysis}

\begin{table}[t]
\centering
\caption{Publication bias and sensitivity analyses for the three educational domains}
\label{tab:pub_bias_sensitivity}
\resizebox{\textwidth}{!}{
\begin{tabular}{lcccccccc}
\toprule
\multirow{2}{*}{\textbf{Domain}} &
\multicolumn{2}{c}{\textbf{Egger's Test}} &
\multicolumn{2}{c}{\textbf{Trim \& Fill}} &
\multicolumn{2}{c}{\textbf{Fail-safe $N$}} &
\multicolumn{2}{c}{\textbf{Leave-one-out 95\% CI Range}} \\
\cmidrule(lr){2-3}\cmidrule(lr){4-5}\cmidrule(lr){6-7}\cmidrule(lr){8-9}
 & $z$ & $p$ & $k_{miss}$ & SE & $N$ & Threshold & CI$_{lb}$ & CI$_{ub}$ \\
\midrule
Qualification 
 & 3.62 & .0003 & 0 & 6.26 & 4583 & 625 & [0.621, 0.651] & [0.844, 0.877] \\

Socialisation 
 & 0.25 & .8038 & 0 & 2.21 & 34 & 65 & [0.354, 0.574] & [1.003, 1.129] \\

Subjectification 
 & 3.15 & .0016 & 0 & 4.23 & 872 & 280 & [0.483, 0.525] & [0.755, 0.817] \\
\bottomrule
\end{tabular}}
\end{table}

\begin{figure}[htbp]
  \centering
  \includegraphics[width=0.6\textwidth]{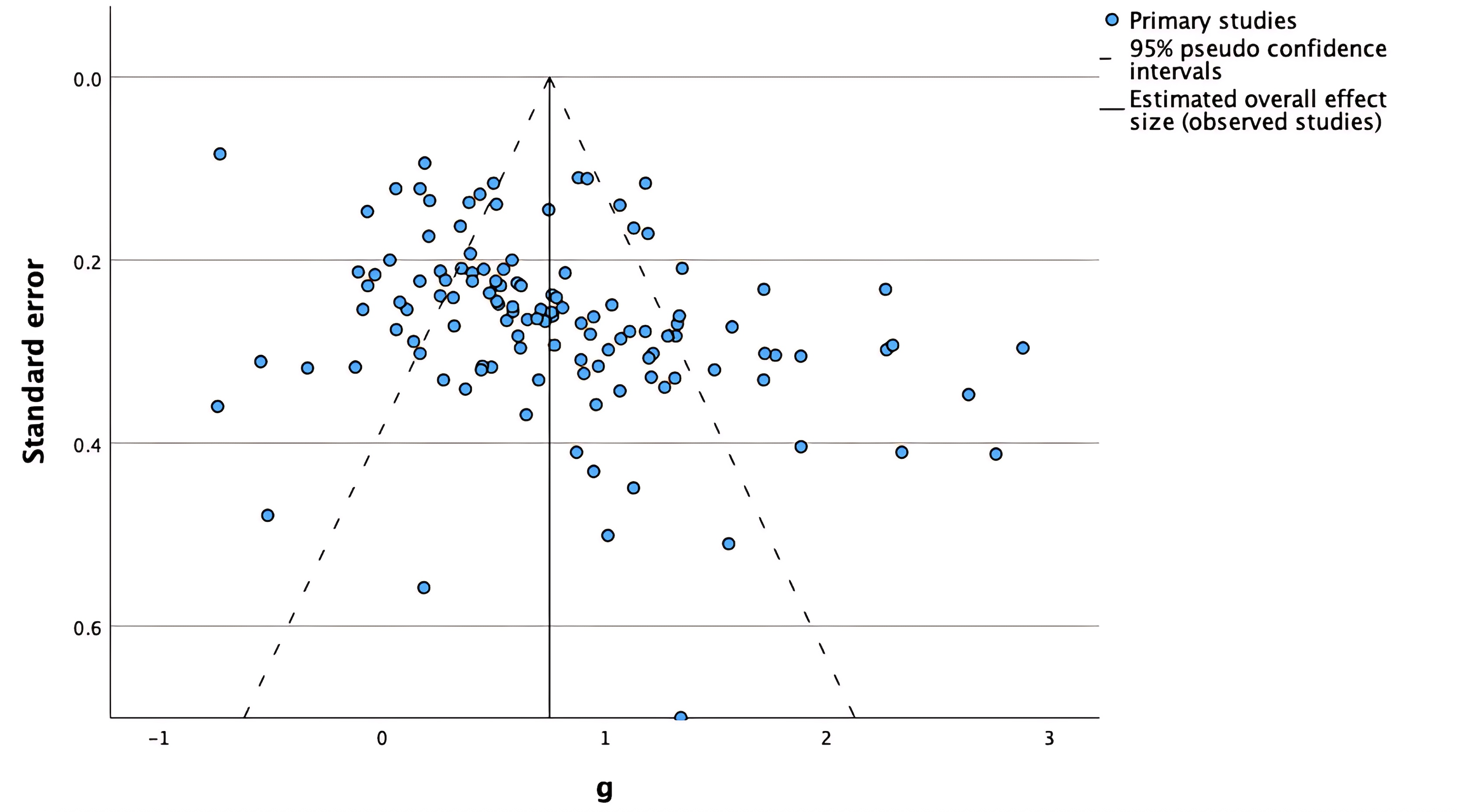}
  \caption{Funnel plot of studies on Qualification (95\% CI).}
  \label{fig:funnel_qualification}
\end{figure}

\begin{figure}[htbp]
  \centering
  \includegraphics[width=0.6\textwidth]{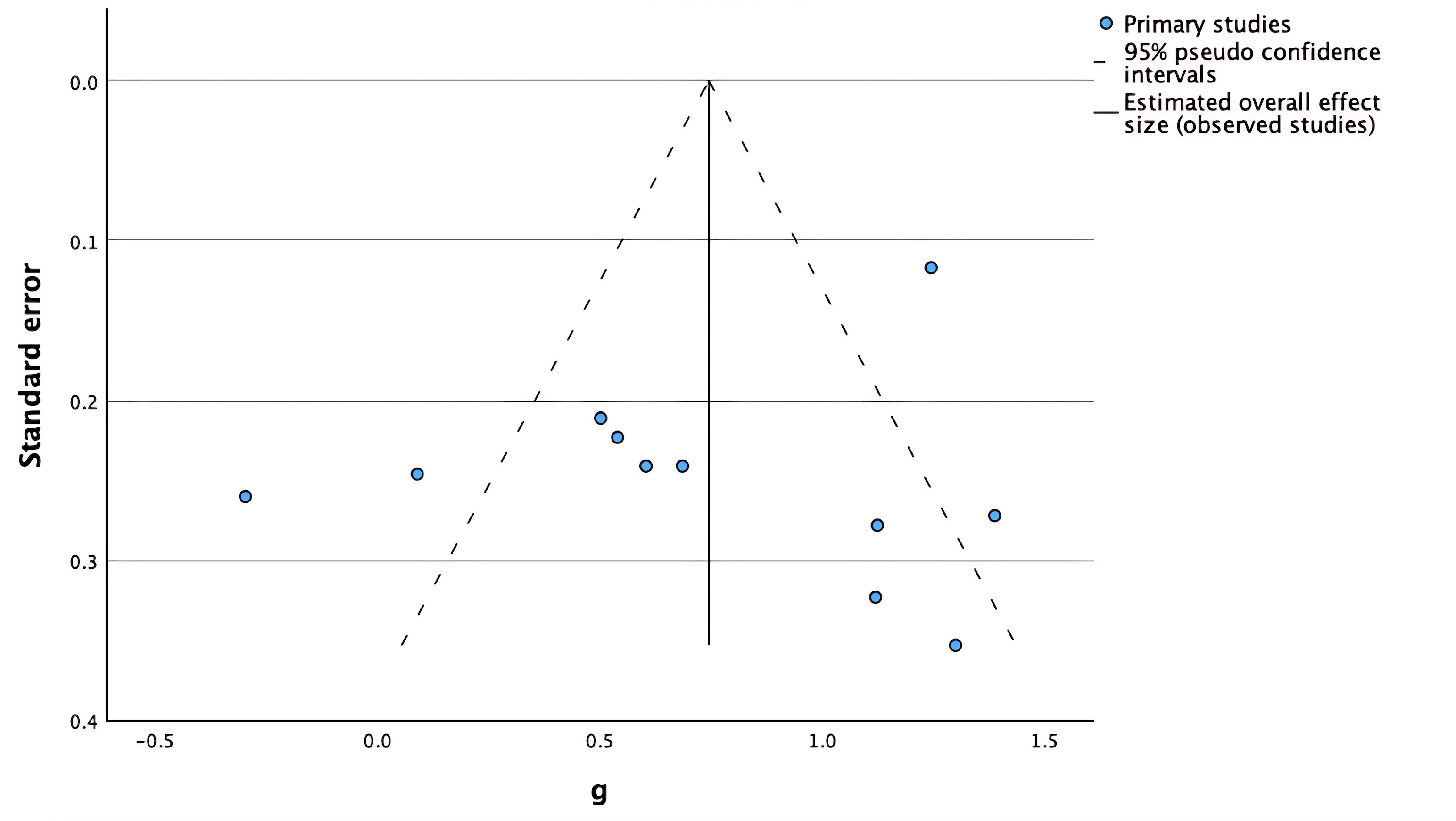}
  \caption{Funnel plot of studies on Socialisation (95\% CI).}
  \label{fig:funnel_socialization}
\end{figure}

\begin{figure}[htbp]
  \centering
  \includegraphics[width=0.6\textwidth]{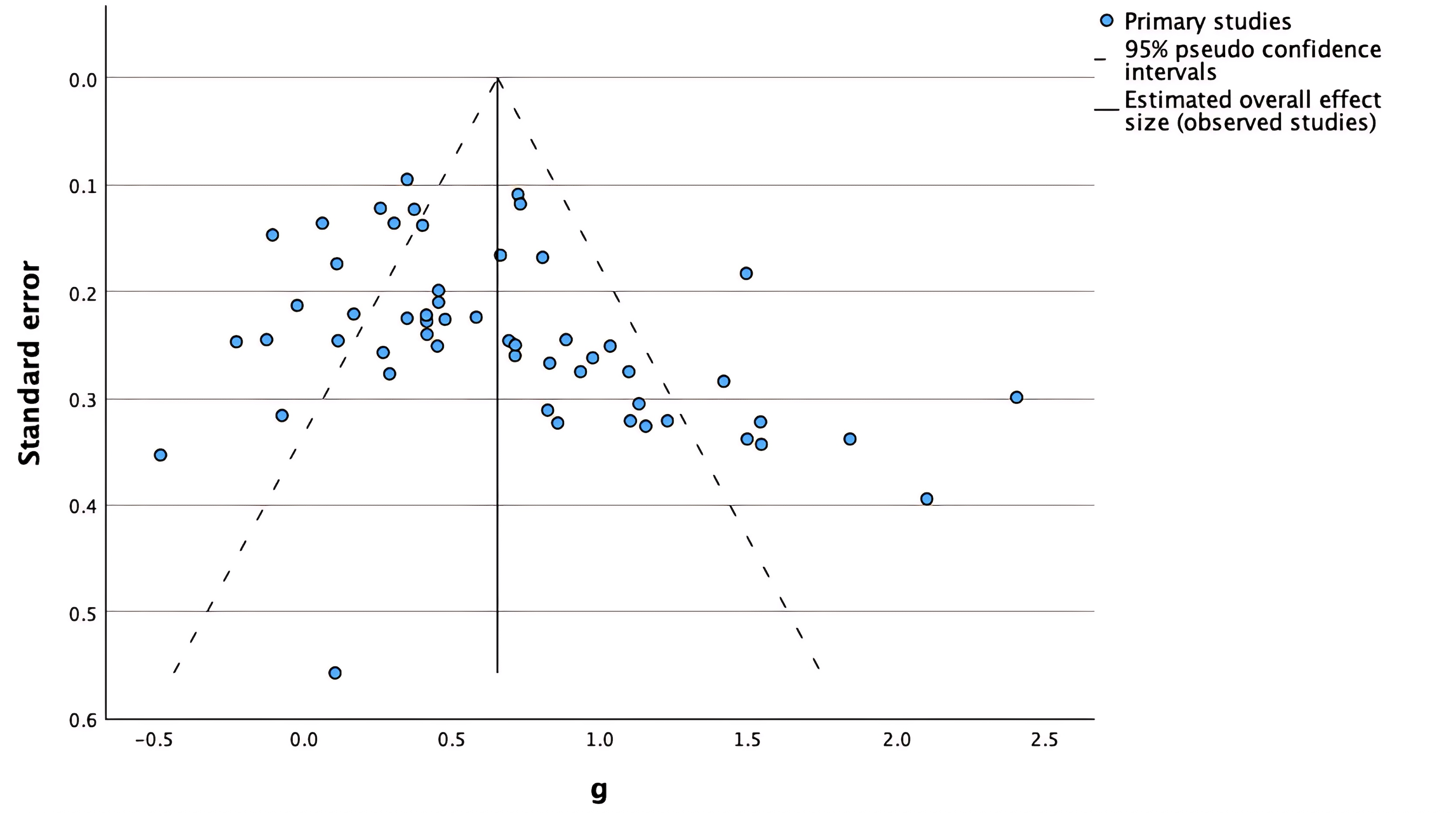}
  \caption{Funnel plot of studies on Subjectification (95\% CI).}
  \label{fig:funnel_subjectification}
\end{figure}

We assessed potential publication bias~\cite{schulz1995empirical} by examining funnel plots for asymmetry, a phenomenon known as small-study effects, where smaller studies tend to report larger effects. Egger's regression test~\cite{egger1997bias} was used to detect such effects statistically.

Visual inspection revealed asymmetry in the funnel plots for Qualification (Figure~\ref{fig:funnel_qualification}) and Subjectification (Figure~\ref{fig:funnel_subjectification}), which Egger's test confirmed as statistically significant ($p<.001$ and $p=.0016$, respectively). In contrast, the plot for Socialisation (Figure~\ref{fig:funnel_socialization}) appeared symmetrical, a finding supported by a non-significant Egger's test ($p=.8038$).

However, further sensitivity analyses indicate that this asymmetry does not threaten our core conclusions; instead, it highlights the varying robustness of the evidence across domains (Table~\ref{tab:pub_bias_sensitivity}). Although small-study effects were present, the Trim and Fill method~\cite{duval2000nonparametric} imputed no missing studies for any domain. More importantly, the findings for Qualification and Subjectification proved highly robust, with Fail-safe $N$s (4583 and 872)~\cite{borenstein2021introduction} massively exceeding their tolerance thresholds. Conversely, the result for Socialisation is fragile, with its Fail-safe $N$ (34) falling below the required threshold of 65, suggesting this finding should be interpreted with caution. Finally, leave-one-out analysis~\cite{viechtbauer2010outlier} confirmed that no single study unduly influenced the overall effect sizes.

\section{Implications for Theory, Method, Practice, and Policy}
\label{sec_implication}

\subsection{Theoretical Implications}
Our study shows that, when we evaluate educational technologies, we need to look beyond simple “does it raise scores?” questions. Using Biesta's framework helps explain why apparently positive results may hide tensions: if we focus mainly on helping students master knowledge and skills (qualification), improvements in collaboration (socialisation) and personal agency (subjectification) become less steady and rely much more on how activities are designed~\cite{biesta2009good,biesta2015good}. Our findings support Biesta's view that developing students' sense of agency is not about removing all structure; rather, it requires well-designed opportunities for them to make choices, explain their reasoning, and take responsibility~\cite{biesta2015beautiful,biesta2021world}. In contexts where large language models can supply answers almost instantly, such structured opportunities may be especially important, as they can help create the conditions under which students' autonomy is more likely to develop.

\subsection{Methodological and Scholarly Implications}
Our analysis tentatively points to several shifts that may help strengthen the methodological and scholarly foundations of research on LLMs in education.

\textit{Adopt a Purpose-Driven Evaluation Lens.}
Future studies may find it useful to report outcomes not only by cognitive taxonomies (e.g., Bloom's) but also by the broader educational purposes: qualification, socialisation, and subjectification. Prior syntheses often emphasise test-based performance, which may make social and personal outcomes less visible~\cite{sun2024does, qu2025generative, liu2025impact}. A purpose-level lens could help surface these dimensions and enable more like-for-like comparisons across designs.

\textit{Assess for Durable, Transferable Learning.}
It may be valuable to pre-register analytic plans and, where feasible, include follow-up or transfer assessments. Because many studies rely on immediate post-tests~\cite{hakiki2023exploring, mackey2024evaluating, sun2024would}, it can be difficult to judge whether observed gains persist or coincide with delayed drawbacks from over-reliance (e.g., weaker performance once AI support is removed). Incorporating delayed measures could help distinguish short-lived boosts from more durable learning~\cite{behforouz2024grammar, sayed2024artificial, farhane2025impact}.

\textit{Broaden the Evidence Base for generalisability.}
Given the current concentration in higher education and language/CS contexts in our sample, generalising to earlier schooling and other subjects remains uncertain. Extending coverage to primary and secondary levels, a wider range of subjects, and under-represented groups may improve transferability and help guard against sample-driven bias in design ~\cite{holmes2023guidance, kasneci2023chatgpt}.

\textit{Demand Granular Reporting of Design Parameters.}
It would be helpful for studies to describe key design choices, for example, the intended role of the LLM (i.e., tutor, collaborator, or tool), prompting/guardrails, duration, and any encouraged study practices (e.g., note-taking)~\cite{bond2024meta, fu2024navigating}. In our moderator analyses, role and duration appeared to be associated with differences in effects. More granular reporting may facilitate replication and support a cumulative account of what works, for whom, for what purpose, and under what conditions.

\subsection{Practice Implications} 
Our findings point to several possible implications for educational practice, offering tentative guidance on how students might engage with LLMs, how teachers could design learning activities, and how institutions might structure their integration. These suggestions are informed by our moderator analysis (RQ3), which indicated that the benefits of LLMs may depend on how they are positioned and scaffolded in learning contexts.

\textit{For Students: Engage as a Thinking Partner, Not an Answer Key.} LLMs may be more effective for qualification when combined with active study habits rather than used as answer providers~\cite{kreijkes2025effects}. Practices such as note-taking, practising recall, and self-explaining ideas have been associated with the deeper cognitive processing required for long-term retention~\cite{dunlosky2013improving, agarwal2024powerful}. Without these strategies, students could risk becoming passive recipients of AI outputs, which might produce superficial short-term gains but undermine durable understanding~\cite{bastani2025generative, darvishi2024impact}. Treating LLMs as thinking partners that complement rather than replace one's own reasoning may help students build transferable knowledge and skills.

\textit{For Teachers: Design for Agency and Transfer.} Our moderator analysis suggested that “tutor-style” prompting appeared to be associated with stronger effects on subjectification. Prompts that invite students to plan, justify, and revise their thinking, while strategically withholding the final answer, may encourage self-regulated learning and reduce over-reliance on AI \cite{pitts2025students, liu2024integrating, xu2025enhancing}. We believe that designing interactions in this way could help shift students from answer-seeking toward agency-driven learning processes, potentially supporting the transfer of learning beyond the immediate task~\cite{steinert2024harnessing, ma2025dbox}.

\textit{For Schools and Courses: Build Scaffolding for Holistic Growth.} At the institutional level, realising the potential of LLMs may benefit from clearer norms and more structured workflows~\cite{spivakovsky2023institutional}. Our analysis suggests that socialisation benefits, such as collaboration and communication, do not typically emerge from isolated tool use but may depend on sustained interventions with time for reflection and peer discussion. Administrators and educators might articulate expected study strategies (e.g., “plan–explain–revise” cycles) and embed these interactional elements into the curriculum to support these social outcomes~\cite{kadaruddin2023empowering, lee2024systematic}. Furthermore, to help ensure that gains in qualification are robust, short follow-up assessments without AI support could be used to check whether students can apply what they have learned independently.

\subsection{Policy Implications}

Policy responses to LLMs in education have often defaulted to polarised positions, either banning them outright or embracing them without restriction \cite{kishore2023should, yu2023reflection}. Such binary framings ignore how design determines impact, risking both the loss of beneficial applications and the unchecked spread of harmful ones. Our results indicate that effective policy should move beyond access decisions and directly regulate design, evaluation, and equity.

\textit{Mandate Design Transparency.}  
Policies should require clear disclosure of the pedagogical role assigned to LLMs (e.g., tutor, collaborator, or tool), the prompting and guardrail strategies used, and the intended learning outcomes. Transparency in these design choices is essential for reproducibility, comparability, and accountability across deployments \cite{memarian2023fairness, chaudhry2022transparency}. Regulators can set minimum reporting standards, similar to clinical trial registries~\cite{ioannidis2005most, sim2006clinical}, that make it harder for providers to market opaque “black-box” educational tools.

\textit{Require Holistic Evaluation.}  
Evaluations should go beyond test scores. Policies ought to require reporting across all three educational purposes: knowledge and skills (qualification), social interaction (socialisation), and personal agency (subjectification), and, where feasible, include a delayed follow-up test~\cite{miao2021artificial}. Such multidimensional assessment ensures that durable learning and broader developmental outcomes are captured, and accreditation bodies or funding agencies can make their approval contingent on its inclusion.

\textit{Direct Funding Toward Equity and Generalisability.}  
Current evidence is heavily skewed toward Asian higher education and language/computer science contexts~\cite{ng2025generative, deng2025does, heung2025chatgpt}. Policies should target funding toward under-represented groups, earlier education levels, and less-studied domains such as arts and mathematics. Support for longer pilot programs in diverse regions will strengthen generalisability and help ensure that global policy debates reflect a wider range of educational realities \cite{pagliara2024integration, li2024systematic, bo2025oecd}.

\textit{Establish Accountability Mechanisms.}  
Finally, policies should embed accountability into the deployment of LLMs in schools and universities. This includes regular audits of actual classroom practices, safeguards against over-reliance, and mechanisms for student and teacher feedback. Without such structures, even well-designed systems risk drifting toward shallow uses that privilege short-term gains over long-term educational purposes.

In summary, effective policy cannot be reduced to whether LLMs are present in classrooms. It should specify how they are designed, evaluated, and distributed. Only then can LLMs contribute reliably to education's full purposes: qualification, socialisation, and subjectification.

\section{Conclusion}
\label{sec_conclusion}

This meta-analysis offers the first large-scale synthesis of how LLMs affect students across Biesta's three purposes of education: qualification, socialisation, and subjectification. Our findings confirm that LLMs reliably enhance qualification, particularly when deployed as sustained, tutor-like interventions. Yet effects on socialisation and subjectification, while positive, are more fragile and context-dependent. These uneven patterns reveal a structural imbalance: LLMs currently strengthen what is most easily measurable, while leaving more relational and developmental purposes vulnerable.

Our findings extend beyond earlier observations that structured use leads to stronger cognitive gains. The moderator analyses presented in this paper identify which design levers matter most: the role assigned to the LLM and the duration of its use, and map their impact onto Biesta’s three educational purposes. This new purpose-level synthesis shows that qualification gains are reliably achieved under tutor-like, sustained designs, while socialisation and subjectification only materialise when participation and agency are deliberately designed into the interaction. Without such intentionality, the broader aims of education remain systematically underdeveloped.

For HCI and educational practice, the implication is practical and normative. Researchers and designers must move beyond treating ``LLM use'' as a single treatment and instead attend to when, how, and for what purposes LLMs are integrated. Educators and policymakers should not equate success with short-term performance alone, but should prioritise designs that scaffold interaction, foster autonomy, and situate learning in community.

Future work should extend evidence beyond higher education and beyond ChatGPT-dominated contexts, incorporate longitudinal and cross-cultural designs, and develop evaluation frameworks that capture not only what students know but also how they belong and who they become. Only by foregrounding these multiple purposes can we realise the promise of LLMs in education without narrowing its horizons.

\bibliographystyle{unsrtnat}
\bibliography{references}  

\begin{thebibliography}{138}
\providecommand{\natexlab}[1]{#1}
\providecommand{\url}[1]{\texttt{#1}}
\expandafter\ifx\csname urlstyle\endcsname\relax
  \providecommand{\doi}[1]{doi: #1}\else
  \providecommand{\doi}{doi: \begingroup \urlstyle{rm}\Url}\fi

\bibitem[Malinka et~al.(2023)Malinka, Peres{\'\i}ni, Firc, Hujn{\'a}k, and Janus]{malinka2023educational}
Kamil Malinka, Martin Peres{\'\i}ni, Anton Firc, Ondrej Hujn{\'a}k, and Filip Janus.
\newblock On the educational impact of chatgpt: Is artificial intelligence ready to obtain a university degree?
\newblock In \emph{Proceedings of the 2023 Conference on Innovation and Technology in Computer Science Education V. 1}, pages 47--53, 2023.

\bibitem[McDonald et~al.(2025)McDonald, Johri, Ali, and Collier]{mcdonald2025generative}
Nora McDonald, Aditya Johri, Areej Ali, and Aayushi~Hingle Collier.
\newblock Generative artificial intelligence in higher education: Evidence from an analysis of institutional policies and guidelines.
\newblock \emph{Computers in Human Behavior: Artificial Humans}, 3:\penalty0 100121, 2025.

\bibitem[Kasneci et~al.(2023)Kasneci, Se{\ss}ler, K{\"u}chemann, Bannert, Dementieva, Fischer, Gasser, Groh, G{\"u}nnemann, H{\"u}llermeier, et~al.]{kasneci2023chatgpt}
Enkelejda Kasneci, Kathrin Se{\ss}ler, Stefan K{\"u}chemann, Maria Bannert, Daryna Dementieva, Frank Fischer, Urs Gasser, Georg Groh, Stephan G{\"u}nnemann, Eyke H{\"u}llermeier, et~al.
\newblock Chatgpt for good? on opportunities and challenges of large language models for education.
\newblock \emph{Learning and individual differences}, 103:\penalty0 102274, 2023.

\bibitem[Bai et~al.(2023)Bai, M{\"u}skens, Zawacki-Richter, and Loglo]{bai2023future}
John~YH Bai, Wolfgang M{\"u}skens, Olaf Zawacki-Richter, and Frank~Senyo Loglo.
\newblock Future prospects of artificial intelligence in education: Developing strategic scenarios to engage educators.
\newblock In \emph{2023: ASCILITE 2023 Conference Proceedings: People, Partnerships and Pedagogies}, pages 22--29, 2023.

\bibitem[Rudolph et~al.(2023)Rudolph, Tan, and Tan]{rudolph2023chatgpt}
J{\"u}rgen Rudolph, Samson Tan, and Shannon Tan.
\newblock Chatgpt: Bullshit spewer or the end of traditional assessments in higher education?
\newblock \emph{Journal of applied learning and teaching}, 6\penalty0 (1):\penalty0 342--363, 2023.

\bibitem[Lehmann et~al.(2024)Lehmann, Cornelius, and Sting]{lehmann2024ai}
Matthias Lehmann, Philipp~B Cornelius, and Fabian~J Sting.
\newblock Ai meets the classroom: When does chatgpt harm learning?
\newblock \emph{Available at SSRN 4941259}, 2024.

\bibitem[Chen et~al.(2024)Chen, Wei, Le, and Zhang]{chen2024learning}
Angxuan Chen, Yuang Wei, Huixiao Le, and Yan Zhang.
\newblock Learning by teaching with chatgpt: The effect of teachable chatgpt agent on programming education.
\newblock \emph{British Journal of Educational Technology}, 2024.

\bibitem[Liu et~al.(2025{\natexlab{a}})Liu, Zhang, and Yang]{liu2025can}
Zhaoyang Liu, Wenlan Zhang, and Panpan Yang.
\newblock Can ai chatbots effectively improve efl learners’ learning effects?—a meta-analysis of empirical research from 2022--2024.
\newblock \emph{Computer Assisted Language Learning}, pages 1--27, 2025{\natexlab{a}}.

\bibitem[Etkin et~al.(2025)Etkin, Etkin, Carter, and Rolle]{etkin2025differential}
Hudson~K Etkin, Kai~J Etkin, Ryan~J Carter, and Camarin~E Rolle.
\newblock Differential effects of gpt-based tools on comprehension of standardized passages.
\newblock In \emph{Frontiers in Education}, volume~10, page 1506752. Frontiers Media SA, 2025.

\bibitem[Bastani et~al.(2025)Bastani, Bastani, Sungu, Ge, Kabakc{\i}, and Mariman]{bastani2025generative}
Hamsa Bastani, Osbert Bastani, Alp Sungu, Haosen Ge, {\"O}zge Kabakc{\i}, and Rei Mariman.
\newblock Generative ai without guardrails can harm learning: Evidence from high school mathematics.
\newblock \emph{Proceedings of the National Academy of Sciences}, 122\penalty0 (26):\penalty0 e2422633122, 2025.

\bibitem[Kreijkes et~al.(2025)Kreijkes, Kewenig, Kuvalja, Lee, Vitello, Hofman, Sellen, Rintel, Goldstein, Rothschild, et~al.]{kreijkes2025effects}
Pia Kreijkes, Viktor Kewenig, Martina Kuvalja, Mina Lee, Sylvia Vitello, Jake~M Hofman, Abigail Sellen, Sean Rintel, Daniel~G Goldstein, David~M Rothschild, et~al.
\newblock Effects of llm use and note-taking on reading comprehension and memory: A randomised experiment in secondary schools.
\newblock \emph{Available at SSRN}, 2025.

\bibitem[Zhai et~al.(2024)Zhai, Wibowo, and Li]{zhai2024effects}
Chunpeng Zhai, Santoso Wibowo, and Lily~D Li.
\newblock The effects of over-reliance on ai dialogue systems on students' cognitive abilities: a systematic review.
\newblock \emph{Smart Learning Environments}, 11\penalty0 (1):\penalty0 28, 2024.

\bibitem[Vidal et~al.(2023)Vidal, Vincent-Lancrin, and Yun]{vidal2023emerging}
Quentin Vidal, St{\'e}phan Vincent-Lancrin, and Hyunkyeong Yun.
\newblock Emerging governance of generative ai in education.
\newblock 2023.

\bibitem[Holmes et~al.(2023)Holmes, Miao, et~al.]{holmes2023guidance}
Wayne Holmes, Fengchun Miao, et~al.
\newblock \emph{Guidance for generative AI in education and research}.
\newblock Unesco Publishing, 2023.

\bibitem[Vincent(2023)]{vincent2023new}
James Vincent.
\newblock New york city schools ban access to chatgpt over fears of cheating and misinformation.
\newblock \emph{The Verge}, 5, 2023.

\bibitem[Deng et~al.(2025)Deng, Jiang, Yu, Lu, and Liu]{deng2025does}
Ruiqi Deng, Maoli Jiang, Xinlu Yu, Yuyan Lu, and Shasha Liu.
\newblock Does chatgpt enhance student learning? a systematic review and meta-analysis of experimental studies.
\newblock \emph{Computers \& Education}, 227:\penalty0 105224, 2025.

\bibitem[Heung and Chiu(2025)]{heung2025chatgpt}
Yuk Mui~Elly Heung and Thomas~KF Chiu.
\newblock How chatgpt impacts student engagement from a systematic review and meta-analysis study.
\newblock \emph{Computers and Education: Artificial Intelligence}, 8:\penalty0 100361, 2025.

\bibitem[Zhu et~al.(2025)Zhu, Liu, and Zhao]{zhu2025exploring}
Yinkun Zhu, Qiwen Liu, and Li~Zhao.
\newblock Exploring the impact of generative artificial intelligence on students’ learning outcomes: A meta-analysis.
\newblock \emph{Education and Information Technologies}, pages 1--29, 2025.

\bibitem[Selwyn(2016)]{selwyn2016technology}
Neil Selwyn.
\newblock \emph{Is technology good for education?}
\newblock John Wiley \& Sons, 2016.

\bibitem[Williamson and Piattoeva(2022)]{williamson2022education}
B~Williamson and N~Piattoeva.
\newblock Education governance and datafication.
\newblock \emph{Education and Information Technologies}, 27:\penalty0 3515--3531, 2022.

\bibitem[Biesta(2009)]{biesta2009good}
Gert Biesta.
\newblock Good education in an age of measurement: On the need to reconnect with the question of purpose in education.
\newblock \emph{Educational Assessment, Evaluation and Accountability (formerly: Journal of Personnel Evaluation in Education)}, 21\penalty0 (1):\penalty0 33--46, 2009.

\bibitem[Biesta(2015{\natexlab{a}})]{biesta2015good}
Gert~JJ Biesta.
\newblock \emph{Good education in an age of measurement: Ethics, politics, democracy}.
\newblock Routledge, 2015{\natexlab{a}}.

\bibitem[Biesta(2021)]{biesta2021world}
Gert Biesta.
\newblock \emph{World-centred education: A view for the present}.
\newblock Routledge, 2021.

\bibitem[Page et~al.(2021)Page, McKenzie, Bossuyt, Boutron, Hoffmann, Mulrow, Shamseer, Tetzlaff, Akl, Brennan, et~al.]{page2021prisma}
Matthew~J Page, Joanne~E McKenzie, Patrick~M Bossuyt, Isabelle Boutron, Tammy~C Hoffmann, Cynthia~D Mulrow, Larissa Shamseer, Jennifer~M Tetzlaff, Elie~A Akl, Sue~E Brennan, et~al.
\newblock The prisma 2020 statement: an updated guideline for reporting systematic reviews.
\newblock \emph{bmj}, 372, 2021.

\bibitem[Liberati et~al.(2009)Liberati, Altman, Tetzlaff, Mulrow, G{\o}tzsche, Ioannidis, Clarke, Devereaux, Kleijnen, and Moher]{liberati2009prisma}
Alessandro Liberati, Douglas~G Altman, Jennifer Tetzlaff, Cynthia Mulrow, Peter~C G{\o}tzsche, John~PA Ioannidis, Mike Clarke, Philip~J Devereaux, Jos Kleijnen, and David Moher.
\newblock The prisma statement for reporting systematic reviews and meta-analyses of studies that evaluate healthcare interventions: explanation and elaboration.
\newblock \emph{Bmj}, 339, 2009.

\bibitem[Meyer et~al.(2024)Meyer, Jansen, Schiller, Liebenow, Steinbach, Horbach, and Fleckenstein]{meyer2024using}
Jennifer Meyer, Thorben Jansen, Ronja Schiller, Lucas~W Liebenow, Marlene Steinbach, Andrea Horbach, and Johanna Fleckenstein.
\newblock Using llms to bring evidence-based feedback into the classroom: Ai-generated feedback increases secondary students’ text revision, motivation, and positive emotions.
\newblock \emph{Computers and Education: Artificial Intelligence}, 6:\penalty0 100199, 2024.

\bibitem[Han and Li(2024)]{han2024exploring}
Jining Han and Mimi Li.
\newblock Exploring chatgpt-supported teacher feedback in the efl context.
\newblock \emph{System}, 126:\penalty0 103502, 2024.

\bibitem[Guo and Wang(2024)]{guo2024resist}
Kai Guo and Deliang Wang.
\newblock To resist it or to embrace it? examining chatgpt’s potential to support teacher feedback in efl writing.
\newblock \emph{Education and Information Technologies}, 29\penalty0 (7):\penalty0 8435--8463, 2024.

\bibitem[Urban et~al.(2024)Urban, D{\v{e}}cht{\v{e}}renko, Lukavsk{\`y}, Hrabalov{\'a}, Svacha, Brom, and Urban]{urban2024chatgpt}
Marek Urban, Filip D{\v{e}}cht{\v{e}}renko, Ji{\v{r}}{\'\i} Lukavsk{\`y}, Veronika Hrabalov{\'a}, Filip Svacha, Cyril Brom, and Kamila Urban.
\newblock Chatgpt improves creative problem-solving performance in university students: An experimental study.
\newblock \emph{Computers \& Education}, 215:\penalty0 105031, 2024.

\bibitem[Sun et~al.(2024)Sun, Boudouaia, Zhu, and Li]{sun2024would}
Dan Sun, Azzeddine Boudouaia, Chengcong Zhu, and Yan Li.
\newblock Would chatgpt-facilitated programming mode impact college students’ programming behaviors, performances, and perceptions? an empirical study.
\newblock \emph{International Journal of Educational Technology in Higher Education}, 21\penalty0 (1):\penalty0 14, 2024.

\bibitem[Yusof(2025)]{yusof2025chatgpt}
Ibnatul~Jalilah Yusof.
\newblock Chatgpt-assisted retrieval practice and exam scores: Does it work?
\newblock \emph{Journal of Information Technology Education: Research}, 24:\penalty0 008, 2025.

\bibitem[Al-Abri(2025)]{al2025exploring}
Abdullah Al-Abri.
\newblock Exploring chatgpt as a virtual tutor: A multi-dimensional analysis of large language models in academic support.
\newblock \emph{Education and Information Technologies}, pages 1--36, 2025.

\bibitem[Deep et~al.(2025)Deep, Martirosyan, Ghosh, and Rahaman]{deep2025chatgpt}
Promethi~Das Deep, Nara Martirosyan, Nitu Ghosh, and Md~Shiblur Rahaman.
\newblock Chatgpt in esl higher education: Enhancing writing, engagement, and learning outcomes.
\newblock \emph{Information}, 16\penalty0 (4):\penalty0 316, 2025.

\bibitem[Baskara et~al.(2023)]{baskara2023exploring}
Risang Baskara et~al.
\newblock Exploring the implications of chatgpt for language learning in higher education.
\newblock \emph{Indonesian Journal of English Language Teaching and Applied Linguistics}, 7\penalty0 (2):\penalty0 343--358, 2023.

\bibitem[Yu et~al.(2025)Yu, Guo, Yang, Zhang, and Dong]{yu2025can}
Hao Yu, Yunyun Guo, Hailiang Yang, Weiyu Zhang, and Yan Dong.
\newblock Can chatgpt revolutionize language learning? unveiling the power of ai in multilingual education through user insights and pedagogical impact.
\newblock \emph{European Journal of Education}, 60\penalty0 (1):\penalty0 e12749, 2025.

\bibitem[Chen et~al.(2023)Chen, Huang, Chen, Tseng, and Li]{chen2023gptutor}
Eason Chen, Ray Huang, Han-Shin Chen, Yuen-Hsien Tseng, and Liang-Yi Li.
\newblock Gptutor: a chatgpt-powered programming tool for code explanation.
\newblock In \emph{International conference on artificial intelligence in education}, pages 321--327. Springer, 2023.

\bibitem[Husain(2024)]{husain2024potentials}
Anas Husain.
\newblock Potentials of chatgpt in computer programming: Insights from programming instructors.
\newblock \emph{Journal of Information Technology Education: Research}, 23:\penalty0 002, 2024.

\bibitem[Mackey et~al.(2024)Mackey, Garabet, Maule, Tadesse, Cross, and Weingarten]{mackey2024evaluating}
Brendan~P Mackey, Razmig Garabet, Laura Maule, Abay Tadesse, James Cross, and Michael Weingarten.
\newblock Evaluating chatgpt-4 in medical education: an assessment of subject exam performance reveals limitations in clinical curriculum support for students.
\newblock \emph{Discover Artificial Intelligence}, 4\penalty0 (1):\penalty0 38, 2024.

\bibitem[Aster et~al.(2025)Aster, Laupichler, Rockwell-Kollmann, Masala, Bala, and Raupach]{aster2025chatgpt}
Alexandra Aster, Matthias~Carl Laupichler, Tamina Rockwell-Kollmann, Gilda Masala, Ebru Bala, and Tobias Raupach.
\newblock Chatgpt and other large language models in medical education—scoping literature review.
\newblock \emph{Medical Science Educator}, 35\penalty0 (1):\penalty0 555--567, 2025.

\bibitem[Vanzo et~al.(2024)Vanzo, Chowdhury, and Sachan]{vanzo2024gpt}
Alessandro Vanzo, Sankalan~Pal Chowdhury, and Mrinmaya Sachan.
\newblock Gpt-4 as a homework tutor can improve student engagement and learning outcomes.
\newblock \emph{arXiv preprint arXiv:2409.15981}, 2024.

\bibitem[Sun and Zhou(2024)]{sun2024does}
Lihui Sun and Liang Zhou.
\newblock Does generative artificial intelligence improve the academic achievement of college students? a meta-analysis.
\newblock \emph{Journal of Educational Computing Research}, 62\penalty0 (7):\penalty0 1676--1713, 2024.

\bibitem[Liu et~al.(2025{\natexlab{b}})Liu, Guo, He, and Hu]{liu2025effects}
Xiaohong Liu, Baoxin Guo, Wei He, and Xiaoyong Hu.
\newblock Effects of generative artificial intelligence on k-12 and higher education students’ learning outcomes: A meta-analysis.
\newblock \emph{Journal of Educational Computing Research}, page 07356331251329185, 2025{\natexlab{b}}.

\bibitem[Hu et~al.(2025{\natexlab{a}})Hu, Pang, and Xing]{hu2025evaluating}
De-Xin Hu, Dan-Dan Pang, and Zhe Xing.
\newblock Evaluating the effects of generative ai on student learning outcomes.
\newblock \emph{Educational Technology \& Society}, 28\penalty0 (3):\penalty0 226--240, 2025{\natexlab{a}}.

\bibitem[Qu et~al.(2025)Qu, Sherwood, Liu, and Aleisa]{qu2025generative}
Xiaodong Qu, Joshua Sherwood, Peiyan Liu, and Nawwaf Aleisa.
\newblock Generative ai tools in higher education: A meta-analysis of cognitive impact.
\newblock In \emph{Proceedings of the Extended Abstracts of the CHI Conference on Human Factors in Computing Systems}, pages 1--9, 2025.

\bibitem[Wang and Fan(2025)]{wang2025effect}
Jin Wang and Wenxiang Fan.
\newblock The effect of chatgpt on students’ learning performance, learning perception, and higher-order thinking: insights from a meta-analysis.
\newblock \emph{Humanities and Social Sciences Communications}, 12\penalty0 (1):\penalty0 1--21, 2025.

\bibitem[Liu et~al.(2025{\natexlab{c}})Liu, Zuo, and Lu]{liu2025impact}
Zhiwei Liu, Haode Zuo, and Yongjing Lu.
\newblock The impact of chatgpt on students' academic achievement: A meta-analysis.
\newblock \emph{Journal of Computer Assisted Learning}, 41\penalty0 (4):\penalty0 e70096, 2025{\natexlab{c}}.

\bibitem[Xia et~al.(2025)Xia, Zhang, Huang, and Chiu]{xia2025impact}
Qi~Xia, Peng Zhang, Wendan Huang, and Thomas~KF Chiu.
\newblock The impact of generative ai on university students’ learning outcomes via bloom’s taxonomy: a meta-analysis and pattern mining approach.
\newblock \emph{Asia Pacific Journal of Education}, pages 1--31, 2025.

\bibitem[Krathwohl(2002)]{krathwohl2002revision}
David~R Krathwohl.
\newblock A revision of bloom's taxonomy: An overview.
\newblock \emph{Theory into practice}, 41\penalty0 (4):\penalty0 212--218, 2002.

\bibitem[Biesta(2024)]{biesta2024taking}
Gert Biesta.
\newblock Taking education seriously: The ongoing challenge.
\newblock \emph{Educational Theory}, 74\penalty0 (3):\penalty0 434--448, 2024.

\bibitem[Knox et~al.(2020)Knox, Williamson, and Bayne]{knox2020machine}
Jeremy Knox, Ben Williamson, and Sian Bayne.
\newblock Machine behaviourism: Future visions of ‘learnification’and ‘datafication’across humans and digital technologies.
\newblock \emph{Learning, Media and Technology}, 45\penalty0 (1):\penalty0 31--45, 2020.

\bibitem[Perrotta and Selwyn(2020)]{perrotta2020deep}
Carlo Perrotta and Neil Selwyn.
\newblock Deep learning goes to school: Toward a relational understanding of ai in education.
\newblock \emph{Learning, media and technology}, 45\penalty0 (3):\penalty0 251--269, 2020.

\bibitem[Coelho et~al.(2025)Coelho, Ham, and Jones]{coelho2025understanding}
Dalila~Pinto Coelho, Miriam Ham, and Sarah-Louise Jones.
\newblock Understanding biesta's three purposes of education: A framework proposal.
\newblock \emph{British Educational Research Journal}, 2025.

\bibitem[Zheng et~al.(2023)Zheng, Niu, Zhong, and Gyasi]{zheng2023effectiveness}
Lanqin Zheng, Jiayu Niu, Lu~Zhong, and Juliana~Fosua Gyasi.
\newblock The effectiveness of artificial intelligence on learning achievement and learning perception: A meta-analysis.
\newblock \emph{Interactive Learning Environments}, 31\penalty0 (9):\penalty0 5650--5664, 2023.

\bibitem[Lin et~al.(2023)Lin, Huang, and Lu]{lin2023artificial}
Chien-Chang Lin, Anna~YQ Huang, and Owen~HT Lu.
\newblock Artificial intelligence in intelligent tutoring systems toward sustainable education: a systematic review.
\newblock \emph{Smart learning environments}, 10\penalty0 (1):\penalty0 41, 2023.

\bibitem[Li et~al.(2024{\natexlab{a}})Li, Ji, and Zhan]{li2024expert}
Tingting Li, Yu~Ji, and Zehui Zhan.
\newblock Expert or machine? comparing the effect of pairing student teacher with in-service teacher and chatgpt on their critical thinking, learning performance, and cognitive load in an integrated-stem course.
\newblock \emph{Asia Pacific Journal of Education}, 44\penalty0 (1):\penalty0 45--60, 2024{\natexlab{a}}.

\bibitem[Pan et~al.(2023)Pan, Qin, Zhang, Lou, Yu, and Yang]{pan2023exploring}
Rouye Pan, Zihan Qin, Lan Zhang, Ligao Lou, Huiju Yu, and Junfeng Yang.
\newblock Exploring the impact of intelligent learning tools on students’ independent learning abilities: a pls-sem analysis of grade 6 students in china.
\newblock \emph{Humanities and Social Sciences Communications}, 10\penalty0 (1):\penalty0 1--11, 2023.

\bibitem[Escalante et~al.(2023)Escalante, Pack, and Barrett]{escalante2023ai}
Juan Escalante, Austin Pack, and Alex Barrett.
\newblock Ai-generated feedback on writing: Insights into efficacy and enl student preference.
\newblock \emph{International Journal of Educational Technology in Higher Education}, 20\penalty0 (1):\penalty0 57, 2023.

\bibitem[Perifanou and Economides(2025)]{perifanou2025collaborative}
Maria Perifanou and Anastasios~A Economides.
\newblock Collaborative uses of genai tools in project-based learning.
\newblock \emph{Education Sciences}, 15\penalty0 (3):\penalty0 354, 2025.

\bibitem[Hamid et~al.(2023)Hamid, Zulkifli, Naimat, Yaacob, and Ng]{hamid2023exploratory}
Hazrina Hamid, Khadjizah Zulkifli, Faiza Naimat, Nor Liana~Che Yaacob, and Kwok~Wen Ng.
\newblock Exploratory study on student perception on the use of chat ai in process-driven problem-based learning.
\newblock \emph{Currents in Pharmacy Teaching and Learning}, 15\penalty0 (12):\penalty0 1017--1025, 2023.

\bibitem[Fu et~al.(2023)Fu, Peng, Khot, and Lapata]{fu2023improving}
Yao Fu, Hao Peng, Tushar Khot, and Mirella Lapata.
\newblock Improving language model negotiation with self-play and in-context learning from ai feedback.
\newblock \emph{arXiv preprint arXiv:2305.10142}, 2023.

\bibitem[Samuel et~al.(2025)Samuel, Soh, and Jung]{samuel2025enhancing}
Anita Samuel, Michael Soh, and Eulho Jung.
\newblock Enhancing reflective practice with chatgpt: A new approach to assignment design.
\newblock \emph{Medical Teacher}, pages 1--3, 2025.

\bibitem[Kassenkhan et~al.(2025)Kassenkhan, Moldagulova, and Serbin]{kassenkhan2025gamification}
Aray~M Kassenkhan, AIMAN~N Moldagulova, and Vasiliy~V Serbin.
\newblock Gamification and artificial intelligence in education: A review of innovative approaches to fostering critical thinking.
\newblock \emph{IEEE Access}, 2025.

\bibitem[Viechtbauer(2010)]{viechtbauer2010conducting}
Wolfgang Viechtbauer.
\newblock Conducting meta-analyses in r with the metafor package.
\newblock \emph{Journal of statistical software}, 36:\penalty0 1--48, 2010.

\bibitem[Hedges(1981)]{hedges1981distribution}
Larry~V Hedges.
\newblock Distribution theory for glass's estimator of effect size and related estimators.
\newblock \emph{journal of Educational Statistics}, 6\penalty0 (2):\penalty0 107--128, 1981.

\bibitem[Lipsey and Wilson(2001)]{lipsey2001practical}
Mark~W Lipsey and David~B Wilson.
\newblock \emph{Practical meta-analysis.}
\newblock SAGE publications, Inc, 2001.

\bibitem[Sawilowsky(2009)]{sawilowsky2009new}
Shlomo~S Sawilowsky.
\newblock New effect size rules of thumb.
\newblock \emph{Journal of modern applied statistical methods}, 8\penalty0 (2):\penalty0 26, 2009.

\bibitem[Zhang et~al.(2024)Zhang, Liang, Tian, and Yu]{zhang2024effects}
Yanjun Zhang, Yanping Liang, Xiaohong Tian, and Xiao Yu.
\newblock The effects of unplugged programming activities on k-9 students’ computational thinking: meta-analysis.
\newblock \emph{Educational technology research and development}, 72\penalty0 (3):\penalty0 1331--1356, 2024.

\bibitem[Higgins et~al.(2003)Higgins, Thompson, Deeks, and Altman]{higgins2003measuring}
Julian~PT Higgins, Simon~G Thompson, Jonathan~J Deeks, and Douglas~G Altman.
\newblock Measuring inconsistency in meta-analyses.
\newblock \emph{bmj}, 327\penalty0 (7414):\penalty0 557--560, 2003.

\bibitem[Kestin et~al.(2025)Kestin, Miller, Klales, Milbourne, and Ponti]{kestin2025ai}
Greg Kestin, Kelly Miller, Anna Klales, Timothy Milbourne, and Gregorio Ponti.
\newblock Ai tutoring outperforms in-class active learning: an rct introducing a novel research-based design in an authentic educational setting.
\newblock \emph{Scientific Reports}, 15\penalty0 (1):\penalty0 17458, 2025.

\bibitem[Scarlatos et~al.(2025)Scarlatos, Liu, Lee, Baraniuk, and Lan]{scarlatos2025training}
Alexander Scarlatos, Naiming Liu, Jaewook Lee, Richard Baraniuk, and Andrew Lan.
\newblock Training llm-based tutors to improve student learning outcomes in dialogues.
\newblock In \emph{International Conference on Artificial Intelligence in Education}, pages 251--266. Springer, 2025.

\bibitem[Noy and Zhang(2023)]{noy2023experimental}
Shakked Noy and Whitney Zhang.
\newblock Experimental evidence on the productivity effects of generative artificial intelligence.
\newblock \emph{Science}, 381\penalty0 (6654):\penalty0 187--192, 2023.

\bibitem[Alneyadi and Wardat(2023)]{alneyadi2023chatgpt}
Saif Alneyadi and Yousef Wardat.
\newblock Chatgpt: Revolutionizing student achievement in the electronic magnetism unit for eleventh-grade students in emirates schools.
\newblock \emph{Contemporary Educational Technology}, 15\penalty0 (4):\penalty0 ep448, 2023.

\bibitem[Kovari(2025)]{kovari2025systematic}
Attila Kovari.
\newblock A systematic review of ai-powered collaborative learning in higher education: Trends and outcomes from the last decade.
\newblock \emph{Social Sciences \& Humanities Open}, 11:\penalty0 101335, 2025.

\bibitem[Hu et~al.(2025{\natexlab{b}})Hu, Tian, and Li]{hu2025enhancing}
Wanqing Hu, Jirong Tian, and Yanyan Li.
\newblock Enhancing student engagement in online collaborative writing through a generative ai-based conversational agent.
\newblock \emph{The Internet and Higher Education}, 65:\penalty0 100979, 2025{\natexlab{b}}.

\bibitem[Wei et~al.(2025)Wei, Wang, Lee, and Liu]{wei2025effects}
Xiaodong Wei, Lei Wang, Lap-Kei Lee, and Ruixue Liu.
\newblock The effects of generative ai on collaborative problem-solving and team creativity performance in digital story creation: an experimental study.
\newblock \emph{International Journal of Educational Technology in Higher Education}, 22\penalty0 (1):\penalty0 23, 2025.

\bibitem[Dai et~al.(2025)Dai, Wen, Jiang, Liu, and Zhang]{dai2025students}
Xusheng Dai, Zhaochun Wen, Jianxiao Jiang, Huiqin Liu, and Yu~Zhang.
\newblock How students use ai feedback matters: Experimental evidence on physics achievement and autonomy.
\newblock \emph{arXiv preprint arXiv:2505.08672}, 2025.

\bibitem[Kumar et~al.(2023)Kumar, Musabirov, Reza, Shi, Wang, Williams, Kuzminykh, and Liut]{kumar2023impact}
Harsh Kumar, Ilya Musabirov, Mohi Reza, Jiakai Shi, Xinyuan Wang, Joseph~Jay Williams, Anastasia Kuzminykh, and Michael Liut.
\newblock Impact of guidance and interaction strategies for llm use on learner performance and perception.
\newblock \emph{arXiv preprint arXiv:2310.13712}, 2023.

\bibitem[Chen et~al.(2025)Chen, Ruan, Ju, Yap, and Wang]{chen2025more}
Xinyue Chen, Kunlin Ruan, Kexin~Phyllis Ju, Nathan Yap, and Xu~Wang.
\newblock More ai assistance reduces cognitive engagement: Examining the ai assistance dilemma in ai-supported note-taking.
\newblock \emph{arXiv preprint arXiv:2509.03392}, 2025.

\bibitem[Alm(2024)]{alm2024exploring}
Antonie Alm.
\newblock Exploring autonomy in the ai wilderness: Learner challenges and choices.
\newblock \emph{Education sciences}, 14\penalty0 (12):\penalty0 1369, 2024.

\bibitem[Belkina et~al.(2025)Belkina, Daniel, Nikolic, Haque, Lyden, Neal, Grundy, and Hassan]{belkina2025implementing}
Marina Belkina, Scott Daniel, Sasha Nikolic, Rezwanul Haque, Sarah Lyden, Peter Neal, Sarah Grundy, and Ghulam~M Hassan.
\newblock Implementing generative ai (genai) in higher education: A systematic review of case studies.
\newblock \emph{Computers and Education: Artificial Intelligence}, page 100407, 2025.

\bibitem[Zawacki-Richter et~al.(2019)Zawacki-Richter, Mar{\'\i}n, Bond, and Gouverneur]{zawacki2019systematic}
Olaf Zawacki-Richter, Victoria~I Mar{\'\i}n, Melissa Bond, and Franziska Gouverneur.
\newblock Systematic review of research on artificial intelligence applications in higher education--where are the educators?
\newblock \emph{International journal of educational technology in higher education}, 16\penalty0 (1):\penalty0 1--27, 2019.

\bibitem[Marzano(2025)]{marzano2025generative}
Daniela Marzano.
\newblock Generative artificial intelligence (gai) in teaching and learning processes at the k-12 level: A systematic review: D. marzano.
\newblock \emph{Technology, Knowledge and Learning}, pages 1--41, 2025.

\bibitem[Mouta et~al.(2025)Mouta, Pinto-Llorente, and Torrecilla-S{\'a}nchez]{mouta2025agency}
Ana Mouta, Ana~Mar{\'\i}a Pinto-Llorente, and Eva~Mar{\'\i}a Torrecilla-S{\'a}nchez.
\newblock “where is agency moving to?”: Exploring the interplay between ai technologies in education and human agency.
\newblock \emph{Digital Society}, 4\penalty0 (2):\penalty0 49, 2025.

\bibitem[Joseph(2025)]{joseph2025rethinking}
Soosy Joseph.
\newblock Rethinking assessment: how ai is changing the way we measure student success?
\newblock \emph{AI \& SOCIETY}, pages 1--3, 2025.

\bibitem[Uanachain and Aouad(2025)]{uanachain2025generative}
Daire Maria~Ni Uanachain and Lila~Ibrahim Aouad.
\newblock Generative ai in education: Rethinking learning, assessment \& student agency for the ai era.
\newblock \emph{Thresholds in Education}, 48\penalty0 (1):\penalty0 111--132, 2025.

\bibitem[Lee and Moore(2024)]{lee2024harnessing}
Sophia~Soomin Lee and Robert~L Moore.
\newblock Harnessing generative ai (genai) for automated feedback in higher education: A systematic review.
\newblock \emph{Online Learning}, 28\penalty0 (3):\penalty0 82--106, 2024.

\bibitem[Yavuz et~al.(2025)Yavuz, {\c{C}}elik, and Yava{\c{s}}~{\c{C}}elik]{yavuz2025utilizing}
Fatih Yavuz, {\"O}zg{\"u}r {\c{C}}elik, and Gamze Yava{\c{s}}~{\c{C}}elik.
\newblock Utilizing large language models for efl essay grading: An examination of reliability and validity in rubric-based assessments.
\newblock \emph{British Journal of Educational Technology}, 56\penalty0 (1):\penalty0 150--166, 2025.

\bibitem[Zhao et~al.(2024)Zhao, Chapman, and Sabet]{zhao2024generative}
Jian Zhao, Elaine Chapman, and Peyman~GP Sabet.
\newblock Generative ai and educational assessments: A systematic review.
\newblock \emph{Education Research and Perspectives}, 51:\penalty0 124--155, 2024.

\bibitem[Msambwa et~al.(2025)Msambwa, Wen, and Daniel]{msambwa2025impact}
Msafiri~Mgambi Msambwa, Zhang Wen, and Kangwa Daniel.
\newblock The impact of ai on the personal and collaborative learning environments in higher education.
\newblock \emph{European Journal of Education}, 60\penalty0 (1):\penalty0 e12909, 2025.

\bibitem[Zha et~al.(2025)Zha, Tang, Gong, and Xu]{zha2025colp}
Siyu Zha, Yuanrong Tang, Jiangtao Gong, and Yingqing Xu.
\newblock Colp: Scaffolding children’s online long-term collaborative learning.
\newblock \emph{International Journal of Human--Computer Interaction}, pages 1--23, 2025.

\bibitem[Wiboolyasarin et~al.(2024)Wiboolyasarin, Wiboolyasarin, Suwanwihok, Jinowat, and Muenjanchoey]{wiboolyasarin2024synergizing}
Watcharapol Wiboolyasarin, Kanokpan Wiboolyasarin, Kanpabhat Suwanwihok, Nattawut Jinowat, and Renu Muenjanchoey.
\newblock Synergizing collaborative writing and ai feedback: An investigation into enhancing l2 writing proficiency in wiki-based environments.
\newblock \emph{Computers and Education: Artificial Intelligence}, 6:\penalty0 100228, 2024.

\bibitem[Roe and Perkins(2024)]{roe2024generative}
Jasper Roe and Mike Perkins.
\newblock Generative ai and agency in education: A critical scoping review and thematic analysis.
\newblock \emph{arXiv preprint arXiv:2411.00631}, 2024.

\bibitem[Van~den Berg and Du~Plessis(2023)]{van2023chatgpt}
Geesje Van~den Berg and Elize Du~Plessis.
\newblock Chatgpt and generative ai: Possibilities for its contribution to lesson planning, critical thinking and openness in teacher education.
\newblock \emph{Education Sciences}, 13\penalty0 (10):\penalty0 998, 2023.

\bibitem[Chi and Wylie(2014)]{chi2014icap}
Michelene~TH Chi and Ruth Wylie.
\newblock The icap framework: Linking cognitive engagement to active learning outcomes.
\newblock \emph{Educational psychologist}, 49\penalty0 (4):\penalty0 219--243, 2014.

\bibitem[Szymanski et~al.(2025)Szymanski, Stamper, Vanden~Abeele, and Verbert]{szymanski2025granular}
Maxwell Szymanski, John Stamper, Vero Vanden~Abeele, and Katrien Verbert.
\newblock Granular feedback: Leveraging domain expertise and explainable ai to effectively steer models.
\newblock In \emph{Proceedings of the 33rd ACM Conference on User Modeling, Adaptation and Personalization}, pages 94--103, 2025.

\bibitem[Lin et~al.(2025)Lin, Lee, Wang, Huang, and Wu]{lin2025enhancing}
Chia-Ju Lin, Hsin-Yu Lee, Wei-Sheng Wang, Yueh-Min Huang, and Ting-Ting Wu.
\newblock Enhancing reflective thinking in stem education through experiential learning: The role of generative ai as a learning aid.
\newblock \emph{Education and Information Technologies}, 30\penalty0 (5):\penalty0 6315--6337, 2025.

\bibitem[Pratama et~al.(2023)Pratama, Sampelolo, and Lura]{pratama2023revolutionizing}
Muh~Putra Pratama, Rigel Sampelolo, and Hans Lura.
\newblock Revolutionizing education: harnessing the power of artificial intelligence for personalized learning.
\newblock \emph{Klasikal: Journal of education, language teaching and science}, 5\penalty0 (2):\penalty0 350--357, 2023.

\bibitem[ZHAO et~al.(2023)ZHAO, LU, DENG, ZHENG, WANG, CHOWDHURY, YUN, CUI, XUCHAO, Zhao, et~al.]{zhao2023beyond}
XUJIANG ZHAO, JIAYING LU, CHENGYUAN DENG, C~ZHENG, JUNXIANG WANG, TANMOY CHOWDHURY, L~YUN, HEJIE CUI, ZHANG XUCHAO, Tianjiao Zhao, et~al.
\newblock Beyond one-model-fits-all: A survey of domain specialization for large language models.
\newblock \emph{arXiv preprint arXiv}, 2305, 2023.

\bibitem[Ling et~al.(2023)Ling, Zhao, Lu, Deng, Zheng, Wang, Chowdhury, Li, Cui, Zhang, et~al.]{ling2023domain}
Chen Ling, Xujiang Zhao, Jiaying Lu, Chengyuan Deng, Can Zheng, Junxiang Wang, Tanmoy Chowdhury, Yun Li, Hejie Cui, Xuchao Zhang, et~al.
\newblock Domain specialization as the key to make large language models disruptive: A comprehensive survey.
\newblock \emph{ACM Computing Surveys}, 2023.

\bibitem[Nopas(2025)]{nopas2025algorithmic}
Dech-siri Nopas.
\newblock Algorithmic learning or learner autonomy? rethinking ai’s role in digital education.
\newblock \emph{Qualitative Research Journal}, 2025.

\bibitem[Grund and Tulis(2020)]{grund2020facilitating}
Christian~Karl Grund and Maria Tulis.
\newblock Facilitating student autonomy in large-scale lectures with audience response systems.
\newblock \emph{Educational Technology Research and Development}, 68\penalty0 (3):\penalty0 975--993, 2020.

\bibitem[Biesta(2015{\natexlab{b}})]{biesta2015beautiful}
Gert~JJ Biesta.
\newblock \emph{Beautiful risk of education}.
\newblock Routledge, 2015{\natexlab{b}}.

\bibitem[Darvishi et~al.(2024)Darvishi, Khosravi, Sadiq, Ga{\v{s}}evi{\'c}, and Siemens]{darvishi2024impact}
Ali Darvishi, Hassan Khosravi, Shazia Sadiq, Dragan Ga{\v{s}}evi{\'c}, and George Siemens.
\newblock Impact of ai assistance on student agency.
\newblock \emph{Computers \& Education}, 210:\penalty0 104967, 2024.

\bibitem[Shukla et~al.(2024)Shukla, Terziyan, and Tiihonen]{shukla2024ai}
Amit~K Shukla, Vagan Terziyan, and Timo Tiihonen.
\newblock Ai as a user of ai: Towards responsible autonomy.
\newblock \emph{Heliyon}, 10\penalty0 (11), 2024.

\bibitem[Smith and Darvas(2017)]{smith2017encouraging}
Victoria~D Smith and Janet~W Darvas.
\newblock Encouraging student autonomy through higher order thinking skills.
\newblock \emph{Journal of Instructional Research}, 6:\penalty0 29--34, 2017.

\bibitem[Ahmadi and Izadpanah(2019)]{ahmadi2019study}
Maryam Ahmadi and Siros Izadpanah.
\newblock The study of relationship between learning autonomy, language anxiety, and thinking style: The case of iranian university students.
\newblock \emph{International Journal of Research in English Education}, 4\penalty0 (2):\penalty0 73--88, 2019.

\bibitem[Schulz et~al.(1995)Schulz, Chalmers, Hayes, and Altman]{schulz1995empirical}
Kenneth~F Schulz, Iain Chalmers, Richard~J Hayes, and Douglas~G Altman.
\newblock Empirical evidence of bias: dimensions of methodological quality associated with estimates of treatment effects in controlled trials.
\newblock \emph{Jama}, 273\penalty0 (5):\penalty0 408--412, 1995.

\bibitem[Egger et~al.(1997)Egger, Smith, Schneider, and Minder]{egger1997bias}
Matthias Egger, George~Davey Smith, Martin Schneider, and Christoph Minder.
\newblock Bias in meta-analysis detected by a simple, graphical test.
\newblock \emph{bmj}, 315\penalty0 (7109):\penalty0 629--634, 1997.

\bibitem[Duval and Tweedie(2000)]{duval2000nonparametric}
Sue Duval and Richard Tweedie.
\newblock A nonparametric “trim and fill” method of accounting for publication bias in meta-analysis.
\newblock \emph{Journal of the american statistical association}, 95\penalty0 (449):\penalty0 89--98, 2000.

\bibitem[Borenstein et~al.(2021)Borenstein, Hedges, Higgins, and Rothstein]{borenstein2021introduction}
Michael Borenstein, Larry~V Hedges, Julian~PT Higgins, and Hannah~R Rothstein.
\newblock \emph{Introduction to meta-analysis}.
\newblock John wiley \& sons, 2021.

\bibitem[Viechtbauer and Cheung(2010)]{viechtbauer2010outlier}
Wolfgang Viechtbauer and Mike W-L Cheung.
\newblock Outlier and influence diagnostics for meta-analysis.
\newblock \emph{Research synthesis methods}, 1\penalty0 (2):\penalty0 112--125, 2010.

\bibitem[Hakiki et~al.(2023)Hakiki, Fadli, Samala, Fricticarani, Dayurni, Rahmadani, Astiti, and Sabir]{hakiki2023exploring}
Muhammad Hakiki, Radinal Fadli, Agariadne~Dwinggo Samala, Ade Fricticarani, Popi Dayurni, Kurniati Rahmadani, Ayu~Dewi Astiti, and Arisman Sabir.
\newblock Exploring the impact of using chat-gpt on student learning outcomes in technology learning: The comprehensive experiment.
\newblock \emph{Advances in Mobile Learning Educational Research}, 3\penalty0 (2):\penalty0 859--872, 2023.

\bibitem[Behforouz and Al~Ghaithi(2024)]{behforouz2024grammar}
Behnam Behforouz and Ali Al~Ghaithi.
\newblock Grammar gains: Transforming efl learning with chatgpt.
\newblock \emph{Educational Process: International Journal}, 13\penalty0 (4):\penalty0 25--41, 2024.

\bibitem[Sayed et~al.(2024)Sayed, Bani~Younes, Alkhayyat, Adhamova, and Teferi]{sayed2024artificial}
Biju~Theruvil Sayed, Zein~Bassam Bani~Younes, Ahmad Alkhayyat, Iroda Adhamova, and Habesha Teferi.
\newblock To be with artificial intelligence in oral test or not to be: a probe into the traces of success in speaking skill, psychological well-being, autonomy, and academic buoyancy.
\newblock \emph{Language Testing in Asia}, 14\penalty0 (1):\penalty0 49, 2024.

\bibitem[Farhane(2025)]{farhane2025impact}
Hamza Farhane.
\newblock The impact of chatgpt as a formative feedback tool on moroccan secondary school students’ paragraph writing skills: A quasi-experimental design.
\newblock \emph{Journal of English Language Teaching and Applied Linguistics}, 7\penalty0 (3):\penalty0 258--275, 2025.

\bibitem[Bond et~al.(2024)Bond, Khosravi, De~Laat, Bergdahl, Negrea, Oxley, Pham, Chong, and Siemens]{bond2024meta}
Melissa Bond, Hassan Khosravi, Maarten De~Laat, Nina Bergdahl, Violeta Negrea, Emily Oxley, Phuong Pham, Sin~Wang Chong, and George Siemens.
\newblock A meta systematic review of artificial intelligence in higher education: A call for increased ethics, collaboration, and rigour.
\newblock \emph{International journal of educational technology in higher education}, 21\penalty0 (1):\penalty0 4, 2024.

\bibitem[Fu and Weng(2024)]{fu2024navigating}
Yao Fu and Zhenjie Weng.
\newblock Navigating the ethical terrain of ai in education: A systematic review on framing responsible human-centered ai practices.
\newblock \emph{Computers and Education: Artificial Intelligence}, 7:\penalty0 100306, 2024.

\bibitem[Dunlosky et~al.(2013)Dunlosky, Rawson, Marsh, Nathan, and Willingham]{dunlosky2013improving}
John Dunlosky, Katherine~A Rawson, Elizabeth~J Marsh, Mitchell~J Nathan, and Daniel~T Willingham.
\newblock Improving students’ learning with effective learning techniques: Promising directions from cognitive and educational psychology.
\newblock \emph{Psychological Science in the Public interest}, 14\penalty0 (1):\penalty0 4--58, 2013.

\bibitem[Agarwal and Bain(2024)]{agarwal2024powerful}
Pooja~K Agarwal and Patrice~M Bain.
\newblock \emph{Powerful teaching: Unleash the science of learning}.
\newblock John Wiley \& Sons, 2024.

\bibitem[Pitts et~al.(2025)Pitts, Rani, Mildort, and Cook]{pitts2025students}
Griffin Pitts, Neha Rani, Weedguet Mildort, and Eva-Marie Cook.
\newblock Students' reliance on ai in higher education: Identifying contributing factors.
\newblock \emph{arXiv preprint arXiv:2506.13845}, 2025.

\bibitem[Liu et~al.(2024)Liu, Hwang, Chen, Chen, and Ye]{liu2024integrating}
Ze-Min Liu, Gwo-Jen Hwang, Chuang-Qi Chen, Xiang-Dong Chen, and Xin-Dong Ye.
\newblock Integrating large language models into efl writing instruction: effects on performance, self-regulated learning strategies, and motivation.
\newblock \emph{Computer Assisted Language Learning}, pages 1--25, 2024.

\bibitem[Xu et~al.(2025)Xu, Qiao, Cheng, Liu, and Zhao]{xu2025enhancing}
Xiaoqing Xu, Lifang Qiao, Nuo Cheng, Hongxia Liu, and Wei Zhao.
\newblock Enhancing self-regulated learning and learning experience in generative ai environments: The critical role of metacognitive support.
\newblock \emph{British Journal of Educational Technology}, 2025.

\bibitem[Steinert et~al.(2024)Steinert, Avila, Ruzika, Kuhn, and K{\"u}chemann]{steinert2024harnessing}
Steffen Steinert, Karina~E Avila, Stefan Ruzika, Jochen Kuhn, and Stefan K{\"u}chemann.
\newblock Harnessing large language models to develop research-based learning assistants for formative feedback.
\newblock \emph{Smart Learning Environments}, 11\penalty0 (1):\penalty0 62, 2024.

\bibitem[Ma et~al.(2025)Ma, Wang, Zhang, Ma, and Wang]{ma2025dbox}
Shuai Ma, Junling Wang, Yuanhao Zhang, Xiaojuan Ma, and April~Yi Wang.
\newblock Dbox: Scaffolding algorithmic programming learning through learner-llm co-decomposition.
\newblock In \emph{Proceedings of the 2025 CHI Conference on Human Factors in Computing Systems}, pages 1--20, 2025.

\bibitem[Spivakovsky et~al.(2023)Spivakovsky, Omelchuk, Kobets, Valko, and Malchykova]{spivakovsky2023institutional}
Oleksan~V Spivakovsky, Serhii~A Omelchuk, Vitaliy~V Kobets, Nataliia~V Valko, and Daria~S Malchykova.
\newblock Institutional policies on artificial intelligence in university learning, teaching and research.
\newblock \emph{Information Technologies and Learning Tools}, 97\penalty0 (5):\penalty0 181, 2023.

\bibitem[Kadaruddin(2023)]{kadaruddin2023empowering}
Kadaruddin Kadaruddin.
\newblock Empowering education through generative ai: Innovative instructional strategies for tomorrow's learners.
\newblock \emph{International Journal of Business, Law, and Education}, 4\penalty0 (2):\penalty0 618--625, 2023.

\bibitem[Lee and Kwon(2024)]{lee2024systematic}
Sang~Joon Lee and Kyungbin Kwon.
\newblock A systematic review of ai education in k-12 classrooms from 2018 to 2023: Topics, strategies, and learning outcomes.
\newblock \emph{Computers and Education: Artificial Intelligence}, 6:\penalty0 100211, 2024.

\bibitem[Kishore et~al.(2023)Kishore, Hong, Nguyen, and Qutab]{kishore2023should}
Shohil Kishore, Yvonne Hong, Andy Nguyen, and Saima Qutab.
\newblock Should chatgpt be banned at schools? organizing visions for generative artificial intelligence (ai) in education.
\newblock \emph{Rising like a Phoenix: Emerging from the Pandemic and Reshaping Human Endeavors with Digital Technologies}, 2023.

\bibitem[Yu(2023)]{yu2023reflection}
Hao Yu.
\newblock Reflection on whether chat gpt should be banned by academia from the perspective of education and teaching.
\newblock \emph{Frontiers in Psychology}, 14:\penalty0 1181712, 2023.

\bibitem[Memarian and Doleck(2023)]{memarian2023fairness}
Bahar Memarian and Tenzin Doleck.
\newblock Fairness, accountability, transparency, and ethics (fate) in artificial intelligence (ai) and higher education: A systematic review.
\newblock \emph{Computers and Education: Artificial Intelligence}, 5:\penalty0 100152, 2023.

\bibitem[Chaudhry et~al.(2022)Chaudhry, Cukurova, and Luckin]{chaudhry2022transparency}
Muhammad~Ali Chaudhry, Mutlu Cukurova, and Rose Luckin.
\newblock A transparency index framework for ai in education.
\newblock In \emph{International conference on artificial intelligence in education}, pages 195--198. Springer, 2022.

\bibitem[Ioannidis(2005)]{ioannidis2005most}
John~PA Ioannidis.
\newblock Why most published research findings are false.
\newblock \emph{PLoS medicine}, 2\penalty0 (8):\penalty0 e124, 2005.

\bibitem[Sim et~al.(2006)Sim, Chan, G{\"u}lmezoglu, Evans, and Pang]{sim2006clinical}
Ida Sim, An-Wen Chan, A~Metin G{\"u}lmezoglu, Tim Evans, and Tikki Pang.
\newblock Clinical trial registration: transparency is the watchword.
\newblock \emph{The Lancet}, 367\penalty0 (9523):\penalty0 1631--1633, 2006.

\bibitem[Miao and Holmes(2021)]{miao2021artificial}
F~Miao and W~Holmes.
\newblock Artificial intelligence and education. guidance for policy-makers.
\newblock 2021.

\bibitem[Ng and Ho(2025)]{ng2025generative}
Sai-Leung Ng and Chih-Chung Ho.
\newblock Generative ai in education: Mapping the research landscape through bibliometric analysis.
\newblock \emph{Information}, 16\penalty0 (8):\penalty0 657, 2025.

\bibitem[Pagliara et~al.(2024)Pagliara, Bonavolonta, Pia, Falchi, Zurru, Fenu, and Mura]{pagliara2024integration}
Silvio~Marcello Pagliara, Gianmarco Bonavolonta, Mariella Pia, Stefania Falchi, Antioco~Luigi Zurru, Gianni Fenu, and Antonello Mura.
\newblock The integration of artificial intelligence in inclusive education: A scoping review.
\newblock \emph{Information}, 15\penalty0 (12):\penalty0 774, 2024.

\bibitem[Li et~al.(2024{\natexlab{b}})Li, Yu, and Zhang]{li2024systematic}
Li~Li, Fengchao Yu, and Enting Zhang.
\newblock A systematic review of learning task design for k-12 ai education: Trends, challenges, and opportunities.
\newblock \emph{Computers and Education: Artificial Intelligence}, 6:\penalty0 100217, 2024{\natexlab{b}}.

\bibitem[Bo(2025)]{bo2025oecd}
Nang Sagawah~Win Bo.
\newblock Oecd digital education outlook 2023: Towards an effective education ecosystem.
\newblock \emph{Hungarian Educational Research Journal}, 15\penalty0 (2):\penalty0 284--289, 2025.

\end{thebibliography}






\end{document}